\documentclass[aps,prx,showpacs,amsmath,amssymb,superscriptaddress,twocolumn,longbibliography,nofootinbib,10pt]{revtex4-2}
\usepackage{amsfonts}
\usepackage{amsmath}
\usepackage{amssymb}
\usepackage{graphicx}
\usepackage{color}
\usepackage[bookmarks=false,linkcolor=blue,urlcolor=blue,colorlinks,citecolor=blue]{hyperref}
\usepackage[english]{babel}
\usepackage[utf8]{inputenc}
\usepackage{url}
\usepackage[titletoc,title]{appendix}
\usepackage{epstopdf}
\usepackage{dsfont}
\usepackage{braket}
\usepackage{mathtools}
\usepackage{tikz}
\usepackage[export]{adjustbox}
\usetikzlibrary{arrows.meta}
\usepackage{hyperref}
\usepackage{comment}
\usepackage{soul}
\hypersetup{
    unicode=false, 
    pdftoolbar=false, 
    pdfmenubar=true, 
    pdffitwindow=false, 
    pdfstartview={}, 
    pdfsubject={}, 
    pdfcreator={}, 
    pdfproducer={}, 
    pdfnewwindow=true, 
    colorlinks=true, 
    linkcolor=black, 
    citecolor=black, 
    filecolor=black, 
    urlcolor=black 
}
\usepackage{mathrsfs}
\usepackage{outlines} 
\newcommand{\edits}[1]{\textcolor{black}{#1}}

\newcommand{\kkk}{\mathbf{k}}

\makeatletter
\newcommand{\ostar}{\mathbin{\mathpalette\make@circled\star}}
\newcommand{\make@circled}[2]{%
  \ooalign{$\m@th#1\smallbigcirc{#1}$\cr\hidewidth$\m@th#1#2$\hidewidth\cr}%
}
\newcommand{\smallbigcirc}[1]{%
  \vcenter{\hbox{\scalebox{0.77778}{$\m@th#1\bigcirc$}}}%
}
\makeatother

\newcommand{\mytitle}{
Superconducting Proximity Effect in Two-Dimensional Hole Gases 
}
\begin{document}
    \title{\mytitle} 
    \author{Serafim S. Babkin}
     \affiliation{Institute of Science and Technology Austria (ISTA), Am Campus 1, 3400 Klosterneuburg, Austria}
     \author{Benjamin Joecker}
      \affiliation{Center for Quantum Devices, Niels Bohr Institute, University of Copenhagen, DK-2100 Copenhagen, Denmark}
      \affiliation{NNF Quantum Computing Programme, Niels Bohr Institute, University of Copenhagen, Denmark}
    \author{Karsten Flensberg}
    \email{flensberg@nbi.ku.dk}
     \affiliation{Center for Quantum Devices, Niels Bohr Institute, University of Copenhagen, DK-2100 Copenhagen, Denmark}
    \author{Maksym Serbyn}
    \email{maksym.serbyn@ista.ac.at}
    \affiliation{Institute of Science and Technology Austria (ISTA), Am Campus 1, 3400 Klosterneuburg, Austria}
    \author{Jeroen Danon}
    \email{jeroen.danon@ntnu.no}
    \affiliation{Department of Physics, Norwegian University of Science and Technology, NO-7491 Trondheim, Norway}

\begin{abstract} 
Technology involving hybrid superconductor--semiconductor materials is a promising avenue for engineering quantum devices for information storage, manipulation, and transmission.
Proximity-induced superconducting correlations are an essential part of such devices.
While the proximity effect in the conduction band of common semiconductors is well understood, its manifestation in confined \emph{hole gases}, realized for instance in germanium, is an active area of research. 
Lower-dimensional hole-based systems, particularly in germanium, are emerging as an attractive platform for a variety of solid-state quantum devices, due to their combination of efficient spin and charge control and long coherence times. The recent experimental realization of the proximity effect in germanium thus calls for a theoretical description that is tailored to hole gases.
In this work, we propose a simple model to describe proximity-induced superconductivity in two-dimensional hole gases, incorporating both the heavy-hole (HH) and light-hole (LH) bands. 
We start from the Luttinger--Kohn model, introduce three parameters that characterize hopping across the superconductor--semiconductor interface, and derive explicit intraband and interband effective pairing terms for the HH and LH bands. 
Unlike previous approaches, our theory provides a quantitative relationship between induced pairings and interface properties.
Restricting our general model to an experimentally relevant case where only the HH band crosses the chemical potential, we predict the coexistence of $s$-wave and $d$-wave singlet pairings, along with triplet-type pairings, and modified Zeeman and Rashba spin--orbit couplings. 
Our results thus present a starting point for theoretical modeling of quantum devices based on proximitized hole gases, fueling further progress in quantum technology.
\end{abstract}

\maketitle

\section{Introduction \label{sec:introduction}}

Hybrid solid-state systems comprised of semiconductors and superconductors have been very actively researched in the last decade. 
This has partially been driven by the search for exotic topological bound states~\cite{NayakReview} that can arise when effective $p$-wave superconductivity is induced in a semiconductor~\cite{LeijnseReview,AguadoReview,SauReview}. 
So far, the engineering of such systems has largely been done using indium-arsenide-- or indium-antimonide--superconductor heterostructures, where the band alignment of the materials is favorable for inducing superconducting correlations in the semiconductor~\cite{LutchynReview,FlensbergReview}. 
A very recent development in this direction is the construction of minimal Kitaev chains~\cite{Tsintzis2024} in both one-dimensional~\cite{Dvir2023} and two-dimensional~\cite{tenHaaf2024} proximitized semiconductors. 
Other examples of important applications of superconductor--semiconductor hybrids include electrically gateable transmons~\cite{Larsen2015,Casparis2018}, Andreev (spin) qubits~\cite{Janvier2015,Hays2021,PitaVidal2023}, Cooper-pair splitters~\cite{Hofstetter2009,Wang2022}, and hybrid circuit quantum electrodynamic devices~\cite{BurkardReview2020}.

Recently, superconductor--semiconductor research has started moving from electron-carrier to hole-carrier semiconductors.
From the experimental side, there has been a growing interest in proximitized germanium-based two-dimensional hole gases, which have already successfully been used to create a variety of quantum devices~\cite{Vigneau2019,Zhuo2023,LeBlanc2023,Valentini2024,Leblanc2024,Sagi2024,Kiyooka2024}.
From an application point of view, such germanium-based devices promise a number of advantages, such as potentially all-electrical spin-qubit control, transparent contacts to superconductors, and isotopic purification to avoid the presence of nuclear spins~\cite{ScappucciReview,Fang2023}.
These advantages are partially related to the more complicated $p$-wave nature of the orbital wave functions in the valence band.
The corresponding additional orbital angular momentum degree of freedom for the carriers can be mixed with their spin degree of freedom by atomic spin--orbit coupling (SOC), strain, and confinement potentials~\cite{winklerBook}.
The result is an intricately mixed spin--orbital band structure, where the four lowest-energy bands with total angular momentum $J=3/2$ touch at the $\Gamma$-point.
In the presence of confinement, e.g., to create lower-dimensional hole gases, the so-called heavy-hole band with angular momentum projection $j_z = \pm 3/2$ and the light-hole band with $j_z = \pm 1/2$ split, typically resulting in a heavy-hole-like ground state and a heavy-hole--light-hole splitting of tens of meV.

So far, most theoretical modeling of proximitized hole systems has been based on the simplifying assumption that the induced pairing is diagonal in the heavy-hole--light-hole basis, i.e., that it only mixes electrons and holes separately within the $j_z = \pm 3/2$ and $j_z = \pm 1/2$ angular momentum bands~\cite{mao2012hole,Maier2014,Laubscher2024a,Laubscher2024b}. However,  a more careful analyses show that in general also interband pairing terms can arise.
Even in the case of a perfect translationally and rotationally invariant interface, spin--orbit-induced band mixing can conspire with the induced pairing in the conduction band of the semiconductor, effectively yielding finite superconducting correlations between the heavy and light holes~\cite{Futterer2011,Moghaddam2014}, although they are strongly suppressed by the energy gap between the conduction and valence bands, which would make the induced superconducting gap much smaller than the parent gap, which is typically not the case in experiment.
Also for the case of interfaces with a lower degree of symmetry, e.g., due to strong confinement in multiple directions, the tunneling between the superconductor and semiconductor does not need to conserve (microscopic) angular momentum, allowing for more direct interband proximity effects~\cite{Adelsberger2023}.

These findings highlight limitations of oversimplified models that neglect interband pairing, suggesting that a more microscopic theoretical approach is needed. Such a model cannot be purely phenomenological, as na\"ively introducing all symmetry-allowed pairings as independent phenomenological parameters would result in a theory with little predictive power. The need for an adequate theory is further amplified by an increasing number of experiments that probe hybrid germanium--superconductor systems~\cite{Hendrickx2018,Ridderbos2018,Aggarwal2021,Tosato2023,Hinderling2024,lakic2024}, and provide high resolution data, that could potentially be used to constrain a few phenomenological parameters of a quantitative model, provided such a model exists. Constructing a theoretical model of the proximity effect, tailored to lower-dimensional hole gases, that has a tractable number of phenomenological parameters is the central objective of our work.  

In order to develop such a model, 
we reexamine the physics of the interface between a hole gas with bands with a $p$-type orbital character and an $s$-wave superconductor with electron bands with $s$-type orbital character. 
We simply assume that the interface breaks enough symmetries to allow for coupling between any of the $p$-wave bands and the superconductor bands. 
This gives a more complicated induced pairing matrix than the one that is usually worked with, allowing for both inter- and intraband pairing in the semiconductor.
One of the main results of our work is the derivation of the general structure of this pairing matrix, as well as explicit expressions for its nonzero elements in terms of three phenomenological cross-interface hopping parameters and parameters of the band structure of the hole gas and the superconductor. 
The resulting effective Bogoliubov--de Gennes (BdG) Hamiltonian for the hole gas, written in the space of heavy-hole and light-hole bands, is expected to be a useful starting point for theories of proximitized hole-based devices in a broad variety of geometries and settings.

To illustrate the power of our effective model and allow for experimental predictions, we project the general Hamiltonian to the low-energy heavy-hole subspace, using a Schrieffer--Wolff transformation, and we take additionally Rashba-type SOC and a finite Zeeman effect into account. 
Singling out the heavy-hole subspace is relevant for relatively strong two-dimensional confinement, which is typically the case for heterostructures used in experiments~\cite{Lodari2019,Lodari2022,Costa2024}. 
We derive explicit expressions for the different types of superconducting order parameters (singlet $s$-wave and $d$-wave, as well as triplet) that can be induced in the heavy-hole band. 
Moreover, we show that an anisotropic renormalization of the effective $g$-factor and spin--orbit coupling strength can arise due to the coupling with superconductor. 

Finally, we use our results to derive a few straightforwardly observable physical consequences from the different types of induced order parameters and from the renormalized Zeeman coupling. 
We show that the density of states (which can be probed by tunneling spectroscopy) acquires several logarithmic singularities, instead of the BCS-type square-root singularities, the energies of which are tunable by a magnetic field.
Moreover, upon increasing the strength of the magnetic field we show the emergence of Bogoliubov Fermi surfaces~\cite{Yuan2018} with a pattern that is qualitatively distinct from the one observed in indium arsenide~\cite{Phan_2022} or a proximitized topological insulator~\cite{Zhu2021}. 
We also demonstrate the proximity-induced modulation of the $g$-tensor in the hole gas, specifically focusing on the Zeeman effect in elliptically confined quantum dots.    

The remainder of this paper is organized as follows. 
In Section~\ref{sec:summary}, we discuss a heuristic description of an interface between the two-dimensional hole gas and a thick superconducting film, allowing us to intuitively justify the form of induced superconducting pairing in the heavy-hole--light-hole space assuming the absence of momentum conservation across the interface. 
In Section~\ref{sec:model}, we present the complete calculation that incorporates in-plane crystal momentum conservation, enabling us to derive the full Green function of the proximitized two-dimensional hole gas. 
We then consider the projection onto the heavy-hole subspace, deriving an effective BdG Hamiltonian that captures the induced pairing, along with the modification of the Zeeman and Rashba spin--orbit couplings, and the deformation of the Fermi surface of the two-dimensional hole gas induced by the coupling to superconductor. 
In Section~\ref{sec:observables}, we then use this Hamiltonian to calculate the density of states and Bogoliubov Fermi surfaces~\cite{Yuan2018} for the two-dimensional heavy-hole gas.
Additionally, we consider quasi-zero-dimensional confinement and calculate the proximity-induced modifications of the $g$-tensor in an elliptic quantum dot. Finally, we conclude in Section~\ref{sec:discuss} with a summary of our results and a discussion of open questions.

\section{Heuristic derivation of induced pairing \label{sec:summary}}

\begin{figure*}[t]
  \includegraphics[width=0.95\linewidth]{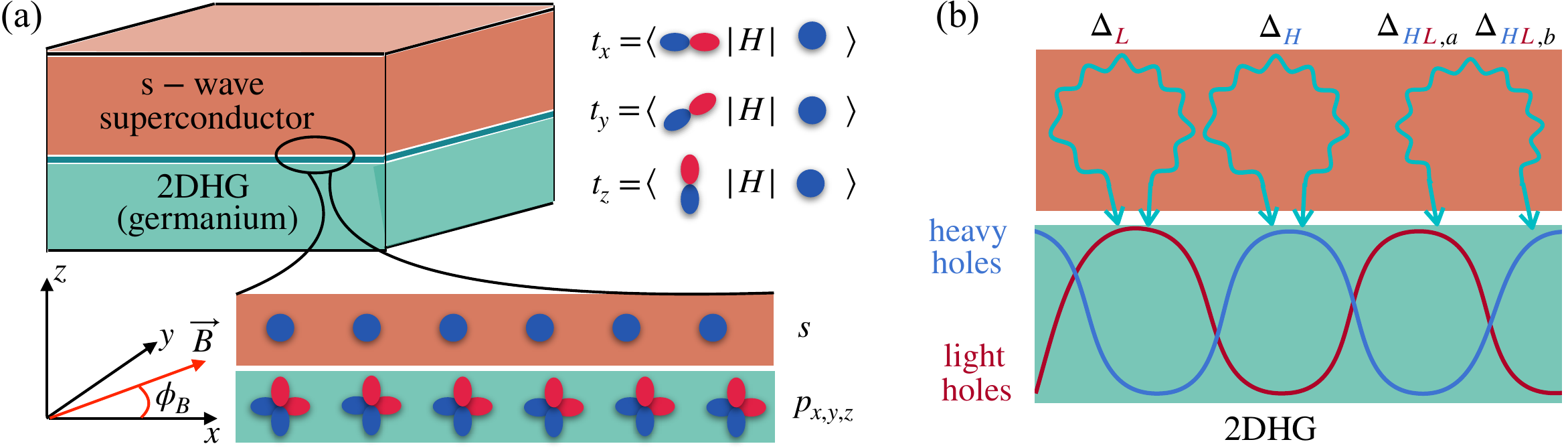}
  \caption{(a) Schematic visualization of the superconductor--semiconductor heterostructure, including the microscopic hopping between the $s$-wave orbitals in the superconductor and the $p$-wave orbitals in the 2DHG.
  We assume that interface breaks rotational symmetry along the $z$-axis, allowing for non-zero values of the phenomenological hopping parameters $t_{x,y,z}$ between the $s$-orbitals and all three $p_{x,y,z}$-orbitals of the 2DHG.
  \edits{Note that the superconductor and semiconductor lattices are drawn as incommensurate, which may act as a symmetry-breaking factor that enhances coupling between $p_{x,y}$- and $s$-orbitals.}
  The potential presence of a spacer between the 2DHG and the superconductor can result in a renormalization of the hopping parameters $t_{x,y,z}$.
  (b) The main result of our work is a microscopic derivation of the proximity-induced pairings between different bands in the 2DHG. Intraband terms $\Delta_{H}$ and $\Delta_{L}$, as well as interband terms $\Delta_{HL,a}$ and $\Delta_{HL,b}$ are illustrated at the top, the blue and red lines at the bottom schematically depict carriers in heavy-hole and light-hole bands of 2DHG, respectively.}\label{fig:cartoon}
\end{figure*}

In this section, we explain the main results and the physical picture of how the band structure of the proximitized two-dimensional hole gas (2DHG) is altered by the induced pairing interaction.

In Fig.~\ref{fig:cartoon}(a) we show schematically the structure we have in mind.
It consists of a germanium (Ge) quantum well in contact with a metallic superconducting top layer.
We mainly focus on a Ge-based 2DHG, but our main results could also be used to describe proximitized hole gases based on other materials. 
Firstly, it is important to consider the nature of the interface between the two materials. 
In some cases the superconductor is in direct contact with the thin Ge layer~\cite{Aggarwal2021,Tosato2023}, while in other cases the two materials are separated by a tunnel barrier of a few nanometers~\cite{Valentini2024}; the physics at the interface is thus not straightforward to model accurately. 
Here, we will assume that the coupling of the Ge hole bands (mostly $p$-orbitals) and the superconductor bands (mostly $s$-orbitals) can be described using spatially independent tunneling amplitudes. 

\begin{figure*}[t]
  \includegraphics[width=0.95\linewidth]{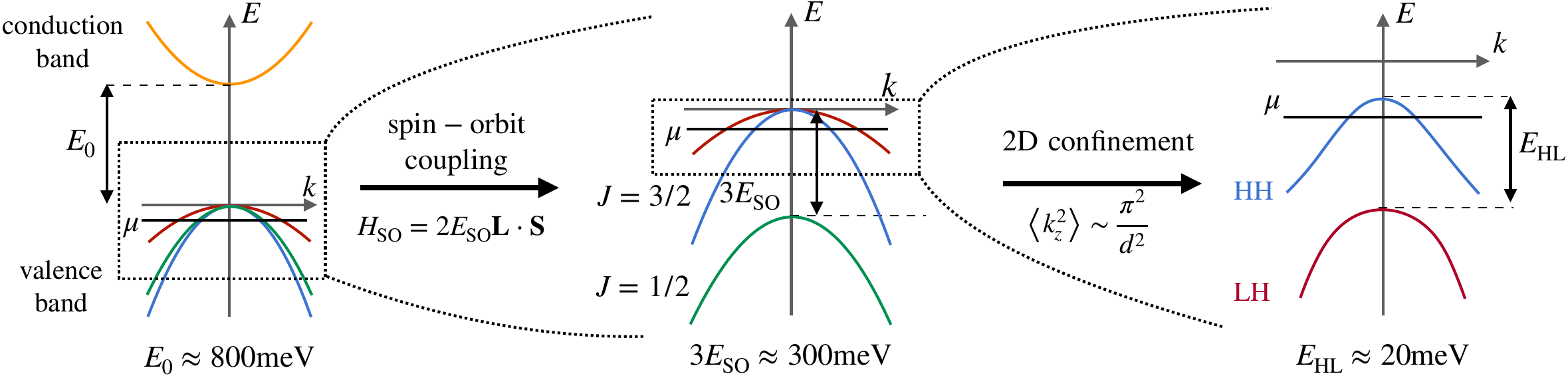}
  \caption{\edits{Schematic visualization of the band structure of a typical semiconductor. The left panel depicts the conduction and valence bands for the 3D case in the absence of spin--orbit coupling. The separation between the bands at $k=0$ is $E_0 \approx 800 \, \mathrm{meV}$, which is much larger than all relevant energy scales in the problem. As a result, the conduction band is disregarded in what follows. The middle panel shows the band structure within the valence band when atomic spin--orbit coupling is taken into account. This coupling results in the splitting of the $J = 3/2$ and $J = 1/2$ bands, with the $J = 1/2$ band being separated by energy $3 E_{\mathrm{SO}} \approx 300 \, \mathrm{meV}$; in what follows, this band is disregarded as well. The right panel depicts the band structure under 2D confinement, resulting in the splitting of the $J = 3/2$ band into heavy-hole (HH) and light-hole (LH) bands, typically separated by $E_{\mathrm{HL}} \approx 20 \, \mathrm{meV}$.}} \label{fig:band_structure}
\end{figure*}

To understand the detailed coupling of the two materials, it is instructive to start from the second-order $\mathbf{k}\cdot\mathbf{p}$ Hamiltonian for the $p$-like valence-band orbitals in the Ge, which reads as~\cite{Luttinger_Kohn,Dresselhaus1955,WillatzenVoonBook}
\begin{widetext}
\begin{equation}\label{HDKK}
	H_{\bf p} = \left(
	\begin{array}{ccc}
		Lk_x^2+M(k_y^2+k_z^2) &Nk_xk_y& Nk_xk_z\\
		Nk_xk_y & Lk_y^2+M(k_x^2+k_z^2)&Nk_yk_z\\
		Nk_xk_z & Nk_yk_z & Lk_z^2+M(k_x^2+k_y^2)
	\end{array}
	\right),
\end{equation}
\end{widetext}
written in the basis $\{ \ket{p_x,\mathbf{k}},\ket{p_y,\mathbf{k}},\ket{p_z,\mathbf{k}} \}$, where $L$, $M$, and $N$ are material-specific parameters (see Fig.\ref{fig:band_structure} for illustration).
The basis states $\ket{p_{x,y,z},\mathbf{k}}$ are Bloch waves where the lattice-periodic part has $p$-orbital symmetry and the wave vectors $\mathbf{k}$ describe the ``envelope'' part of the wave function.

We now focus on the Andreev processes where a particle from the valence band of Ge virtually tunnels above the energy gap in the superconductor and returns as an anti-particle with opposite spin, such as illustrated in Fig.~\ref{fig:cartoon}.
Using second-order perturbation theory, we arrive at the following form of the resulting pairing coupling,
\begin{equation}\label{Delta_ij}
	\Delta_{p_i,p_j} = \sum_{\bf k} \frac{t_{i} t_{j} u_{\bf k} v_{\bf k}}{E_{\bf k}},
\end{equation}
where $u_{\bf k}$ and $v_{\bf k}$ are the particle and anti-particle components of the quasiparticle with momentum $\mathbf{k}$ in the superconductor, $E_{\bf k}$ is the superconductor quasiparticle energy, and $i,j \in \{x,y,z\}$. The parameters $t_i$ describe the tunneling coupling between a state in the semiconductor with $p_i$-wave orbital symmetry and a state in the ($s$-wave) band in the superconductor, see Fig.~\ref{fig:cartoon}(a).
Because the interface generally breaks symmetries of both lattices on the microscopic level, we assume that there are no selection rules that forbid any of the three couplings. 
However, depending on details of the interface, one could find reasons why some of the coupling parameters (e.g., the ``in-plane'' couplings $t_{x,y}$) should be suppressed. We will, however, not include any such assumptions and simply allow all three phenomenological coupling parameters to be non-zero.
Hence, virtual processes in and out of the superconductor can produce couplings that both conserve and do not conserve atomic orbital angular momentum.
We further consider only electrons with an energy close to the Fermi energy and far below the energy gap of the superconductor, which is why the energy of the electron does not appear in the denominator in Eq.~(\ref{Delta_ij}).
As mentioned, the tunneling amplitudes $t_i$ are taken to be spatially constant and the Andreev reflection to be point-like. 
The summation over $\mathbf{k}$ in Eq.~\eqref{Delta_ij} should then be restricted by in-plane momentum conservation. 
However, for simplicity we include all momenta in the summation, which is relevant for a strongly disordered metallic film. We also note that the same result is reached for a thick film where the summation over perpendicular momenta is free. In the more microscopic consideration presented in Sec.~\ref{sec:model}, we will include the dependence of the virtual energy on the state of the electron tunneling into the superconductor, and we will assume the interface to be clean and translationally invariant on the scale of the typical wavelengths in the semiconductor.

Assuming a BCS description of the superconductor, we use the relations $u_{\bf k}v_{\bf k}=\Delta/2E_{\bf k}$  and $E_{\bf k}=\sqrt{\xi_{k}^2+\Delta^2}$ (where $\Delta$ is the gap and $\xi_{k,s} = \hbar^2 k^2 / 2m_s - \mu_s$, with $m_s$ and $\mu_s$ the effective mass and chemical potential of the electrons in the superconductor, respectively) and assuming the normal-state density of states being constant, we can perform the sum in Eq.~\eqref{Delta_ij} by converting it into an integral. The assumption of a constant density of states, $\rho(\xi_{k})\approx \nu_s$ is a simplification which ignores the detailed dependence on the transverse modes (later in the paper we will consider the general case in detail).
We thus arrive at the following approximate proximity-induced pairing amplitude:
\begin{equation}\label{Delta_ij2}
 	\Delta_{p_i,p_j} = \frac{\pi}{2} t_{i} t_{j} \nu_s,
\end{equation}
which does not depend on the gap of the parent superconductor (at energies much smaller than $\Delta$). 

The Andreev reflections described above give rise to an effective induced superconducting pairing in the semiconductor.
To incorporate this effect in the semiconductor Hamiltonian, we include spin in our basis,
enlarge it into Nambu space by applying time reversal, and construct the vector consisting of 12 second-quantized operators
\begin{align}\label{basisspin}
\psi = ( {} & {} d_{p_x,\kkk,\uparrow}, d_{p_x,\kkk,\downarrow}, d_{p_y,\kkk,\uparrow},  d_{p_y,\kkk,\downarrow}, \nonumber\\ {} & {}
 d_{p_z,\kkk,\uparrow},  d_{p_z,\kkk,\downarrow},  d^\dag_{p_x,-\kkk,\downarrow}, -d^\dag_{p_x,-\kkk,\uparrow},  \nonumber\\ {} & {} d^\dag_{p_y,-\kkk,\downarrow}, -d^\dag_{p_y,-\kkk,\uparrow}, d^\dag_{p_z,-\kkk,\downarrow}, -d^\dag_{p_z,-\kkk,\uparrow} )^T
\end{align}
where $d_{p_i,\kkk,\sigma}$ annihilates an electron with spin $\sigma$ in the state $\ket{p_i,\kkk}$.
In this basis, the semiconductor Hamiltonian, including the leading-order proximity effect, can be written in BdG form as
\begin{equation}\label{H0nambu}
	H_0 = \frac{1}{2} \sum_{\bf k} \psi^\dagger\left[ \left(
	\begin{array}{cc}
		H_{\bf p} &\boldsymbol{\Delta}  \\
	\boldsymbol{\Delta}^\dagger & -\Theta 	H_\mathbf{p}\Theta^{-1}
	\end{array}
	\right) \otimes \sigma_0 \right] \psi,
\end{equation}
where $\sigma_i$, with $i=0,x,y,z$, are the Pauli matrices operating in spin space and the elements of the $3\times 3$ matrix $\boldsymbol{\Delta}$ are given by Eq.~\eqref{Delta_ij2}. 

The final ingredient needed to be included is the atomic SOC, which rearranges the $p$-bands in the semiconductor.
In the basis $\{\ket{p_x,\kkk},\ket{p_y,\kkk},\ket{p_z,\kkk}\}\otimes \{\uparrow,\downarrow\}$ the spin--orbit Hamiltonian reads as
\begin{equation}\label{HSO}
	H_{\mathrm{SO}} = 2E_{\mathrm{SO}} \mathbf{L}\cdot\mathbf{S} = E_{\mathrm{SO}} \left(
	\begin{array}{cccccc}
		0&0&-i&0&0&1\\
    	0&0&0&i&-1&0\\
		i&0&0&0&0&-i\\
		0&-i&0&0&-i&0\\
		0&-1&0&i&0&0\\
		1&0&i&0&0&0
	\end{array}
	\right), 
\end{equation}
where ${\bf L}$ and ${\bf S}$ are the orbital and spin angular momentum operators, respectively.
This spin--orbit Hamiltonian can be diagonalized by going to the usual eigenbasis of total angular momentum $\mathbf{J}=\mathbf{L}+\mathbf{S}$, which has two multiplets in this case: one with $J=3/2$ and one with $J=1/2$.
This transforms $H_{\mathrm{SO}}$ to a diagonal form,
\begin{equation}\label{HSOtrans}
	H_{\mathrm{SO}} \longrightarrow E_{\mathrm{SO}} \left(
	\begin{array}{cccccc}
		1&0&0&0&0&0\\
		0&1&0&0&0&0\\
		0&0&1&0&0&0\\
		0&0&0&1&0&0\\
		0&0&0&0&-2&0\\
		0&0&0&0&0&-2
	\end{array}
	\right),
\end{equation} 
which is now written in the basis $
\{\ket{\mathbf{ k},\frac{3}{2},\frac{3}{2}}$, 
$\ket{\mathbf{ k},\frac{3}{2},-\frac{3}{2}}$, 
$\ket{\mathbf{ k},\frac{3}{2},\frac{1}{2}}$, 
$\ket{\mathbf{ k},\frac{3}{2},-\frac{1}{2}}$, 
$\ket{\mathbf{ k},\frac{1}{2},\frac{1}{2}}$, 
$\ket{\mathbf{ k},\frac{1}{2},-\frac{1}{2}}\}$,
using the notation $\ket{{\bf k},J,j_z}$. 
In this basis, the valence-band Hamiltonian $H_{\bf p}\otimes\sigma_0+H_{\rm SO}$ takes the form of the well-known $6\times 6$ Kane model~\cite{winklerBook}.
When ignoring the so-called spin--orbit split-off band with $J=1/2$, which is separated by $3E_{\mathrm{SO}}$, the low-energy Hamiltonian reduces further to a $4\times 4$ Luttinger--Kohn Hamiltonian, describing the light holes (LHs) with $j_z = \pm 1/2$ and heavy holes (HHs) with $j_z = \pm 3/2$ as projection of the total angular momentum $J=3/2$, respectively. 

The important question is how the superconducting pairing matrix $\boldsymbol{\Delta}$ transforms when going to 
the basis that diagonalizes the atomic spin--orbit coupling. In the full basis written in terms of the three $p$-orbitals, the pairing matrix given in Eq.~\eqref{Delta_ij} reads explicitly as
\begin{equation}\label{pairLHHH}
	\boldsymbol{\Delta}\otimes\sigma_0 =
	\left(
	\begin{array}{cccccc}
		\Delta_{xx}&0&\Delta_{xy}&0&\Delta_{xz}&0\\
		0&\Delta_{xx}&0&\Delta_{xy}&0&\Delta_{xz}\\
		\Delta_{xy}&0&\Delta_{yy}&0&\Delta_{yz}&0\\
            0&\Delta_{xy}&0&\Delta_{yy}&0&\Delta_{yz}\\
		\Delta_{xz}&0&\Delta_{yz}&0&\Delta_{zz}&0\\
            0&\Delta_{xz}&0&\Delta_{yz}&0&\Delta_{zz}
	\end{array}
	\right).
\end{equation} 
We perform the rotation to the $\ket{{\bf k},J,j_z}$ basis and restrict the basis to contain 8 second-quantized operators corresponding to LH and HH subspaces only,
\begin{equation} \label{time_rev_basis}
\begin{aligned}
( & d_{\kkk,3/2},d_{\kkk,-3/2},d_{\kkk,1/2},d_{\kkk,-1/2},\\
& -d^\dag_{-\kkk,-3/2},d^\dag_{-\kkk,3/2},d^\dag_{-\kkk,-1/2},-d^\dag_{-\kkk,1/2})^T,
\end{aligned}
\end{equation}
where $d_{{\bf k},j_z}$ annihilates an electron in the state $\ket{{\bf k},\frac{3}{2},j_z}$.
The particular choice of signs on the second line results from the different action of the time-reversal transformation on states with $j_z=\pm 3/2$ and $j_z=\pm 1/2$~\cite{winklerBook}.

In the new basis~(\ref{time_rev_basis}), the pairing matrix acquires the following structure:
\begin{equation}\label{pairLHHHtrans}
		\boldsymbol{\Delta}\otimes\sigma_0
\longrightarrow 
	\left(
	\begin{array}{cccc}		\Delta_{H}&0&\Delta_{HL,a}&\Delta_{HL,b}\\
	0&\Delta_{H}&\Delta_{HL,b}^*&-\Delta_{HL,a}^*\\
		\Delta_{HL,a}^*&\Delta_{HL,b}&\Delta_{L}&0\\
		\Delta_{HL,b}^*&-\Delta_{HL,a}&0&\Delta_{L}
	\end{array}
	\right).
\end{equation} 
Pairings between states that are time-reversed partners, with opposite $j_z$  projection, corresponding to the elements $\Delta_{L}$ and $\Delta_{H}$ in the matrix~(\ref{pairLHHHtrans}) (such pairings lead to terms $\Delta_{L} d^\dagger_{\mathbf{k},1/2}d^\dagger_{-\mathbf{k},-1/2}$ in the BdG Hamiltonian, and similar for $j_z=\pm 3/2$), seem natural at first sight, see Fig.~\ref{fig:cartoon}(b). 
On the other hand, pairings between carriers with different magnitude of $|j_z|$, corresponding to terms like $\Delta_{HL,a} d^\dagger_{\mathbf{k},3/2} d^\dagger_{-\mathbf{k},-1/2}$ and $\Delta_{HL,b} d^\dagger_{\mathbf{k},3/2} d^\dagger_{-\mathbf{k},1/2}$ may seem counterintuitive.
However, the latter are made possible by the singlet proximity-induced pairing because the states $j_z=\pm 1/2$ have both spin up and spin down components. 
Finally, the same-spin intraband pairings vanish due to destructive interference. 

In summary, we see that both intra- and interband pairings can occur, which is one of the main conceptual results of this paper that will be derived in the next section using a more elaborate technique. 
\edits{Qualitatively, this result is largely material-independent, as it relies only on the presence of strong atomic spin--orbit coupling in the semiconductor and finite mixing between all orbitals across the interface. Quantitatively, we}
note that the pairing potentials $\Delta_{H}, \Delta_{L}, \Delta_{HL,a}$ and $\Delta_{HL,b}$ are not arbitrary complex numbers, but instead their relative magnitude is fixed by three real tunnel couplings, via Eq.~(\ref{Delta_ij2}) and a basis rotation. 
This leads to a drastic reduction of the number of phenomenological parameters present in our model. 
We do not provide explicit expressions for the pairings here, since we derive the induced pairing non-perturbatively in hopping in Section~\ref{sec:model} below, which constitutes our main result and is consistent with Eq.~(\ref{pairLHHHtrans}).

\section{Derivation of an effective BdG Hamiltonian for the 2DHG\label{sec:model}}

This section contains the derivation of the effective BdG Hamiltonian for the 2DHG. 
We begin in Sec.~\ref{sec:ge} with the definition of the Hamiltonians describing the semiconductor, the superconductor, and the coupling between them.
In Sec.~\ref{sec:3/2} we then derive the self-energy of the electrons in the semiconductor and obtain an effective Hamiltonian describing the proximity effect in the space of the heavy-hole and light-hole bands, with the same structure as anticipated in Eq.~(\ref{pairLHHHtrans}).
We then project the general BdG Hamiltonian for the 2DHG to the heavy-hole band only, which reveals various effective pairing channels induced by the proximity effect.
We finally discuss their interplay with the effects of Rashba SOC and the Zeeman effect of an applied in-plane magnetic field.

\subsection{Model of the heterostructure\label{sec:ge}}

\subsubsection{Two-dimensional hole gas}
\label{sec:2dhg}

Here, we outline the model we use to describe the electrons in the valence band of the semiconductor.
We introduce the Luttinger--Kohn Hamiltonian, paying special attention to the appropriate basis of the wave functions, since that will be important later when discussing the microscopic tunneling amplitudes between the semiconductor and the superconductor.

As discussed in Sec.~\ref{sec:summary}, the atomic SOC described by $H_{\rm SO}$ mixes the spin of the electrons in the valence band with their orbital angular momentum.
The resulting eigenstates with total angular momentum $J=1/2$ are split off by the spin--orbit energy $3E_{\rm SO} \approx 300~{\rm meV}$ (in Ge) and can therefore be ignored.
The remaining states have total angular momentum $J=3/2$ and can thus be written as $\ket{{\bf k},j_z}$, where $j_z = \pm 3/2, \pm 1/2$ is the total projected angular momentum along $z$.
These states can be written explicitly as
\begin{subequations}
\label{basis}
\begin{align}
&\ket{{\bf k},\tfrac{3}{2}} 
=
-\frac{1}{\sqrt{2}}\big( \ket{p_x,{\bf k},\uparrow} + i \ket{p_y,{\bf k},\uparrow} \big),
\\
&\ket{{\bf k},-\tfrac{3}{2}} 
=
\frac{1}{\sqrt{2}}\big( \ket{p_x,{\bf k},\downarrow} - i \ket{p_y,{\bf k},\downarrow} \big),
\\
&\ket{{\bf k},\tfrac{1}{2}} 
=
\frac{1}{\sqrt{6}}\big( 2\ket{p_z,{\bf k},\uparrow} - \ket{p_x,{\bf k},\downarrow} - i \ket{p_y,{\bf k},\downarrow} \big),
\\
&\ket{{\bf k},-\tfrac{1}{2}} 
=
\frac{1}{\sqrt{6}}\big( 2\ket{p_z,{\bf k},\downarrow} + \ket{p_x,{\bf k},\uparrow} - i \ket{p_y,{\bf k},\uparrow} \big),
\end{align}
\end{subequations}
using the same notation as before.
The valence-band Hamiltonian (\ref{HDKK}) (including spin) projected to the $J=3/2$ subspace and written in this four-dimensional basis is the so-called Luttinger--Kohn Hamiltonian \cite{Luttinger_Kohn,Luttinger_1956}, which reads explicitly as
\begin{align}
    H_\text{LK}= {} & {} \frac{k^2}{2m}\left(\gamma_1+\frac{5}{2}\gamma_2\right) - \frac{\gamma_2}{m}\left(k_x^2J_x^2 + \text{c.p.} \right)
    \nonumber\\&
    - \frac{2\gamma_3}{m}\big( \{k_x,k_y\}\left\{J_x,J_y\right\} + \text{c.p.} \big),
    \label{eq:bulk_Luttinger}
\end{align}
where we have set $\hbar=1$ and used the symmetrization $\{A,B\}=\frac{1}{2}\left(AB+BA\right)$.
Further, $m$ is the electron rest mass (which we formally take to be negative, to capture the negative dispersion of the valence band), $J_{i}$ are the three spin-$3/2$ matrices, and c.p.~denotes cyclic permutation, $x\to y \to z$.
The dimensionless constants $\gamma_{1,2,3}$ are the three material-dependent Luttinger parameters.
In many semiconductors (including GaAs, Ge, and InAs) the difference $|\gamma_2 - \gamma_3|$ is smaller than both $\gamma_2 + \gamma_3$ and $\gamma_1 + \frac{5}{2}\gamma_2$~\cite{winklerBook}, which allows us to use the approximation $\gamma_2 = \gamma_3 \equiv \gamma_s$, yielding a spherically symmetric Hamiltonian.

We now take into account the confinement imposed along the $z$-axis, used to create an effectively two-dimensional hole gas.
This confinement leads to a quantization of the $k_{z}$ component of momentum that is determined by the thickness $d$ of the quantum well.
In order to obtain a Hamiltonian for the 2DHG that depends only on the in-plane  momentum components $k_{x,y}$, we thus project the Hamiltonian on the lowest transverse band by replacing $k_z$ and $k_z^2$ by their expectation values,   
$\left\langle k_{z}^{2}\right\rangle \sim {\pi^{2}}/{d^{2}}$ and $\left\langle k_{z}\right\rangle =0$.
This projection results in the spitting of heavy-hole and light-hole bands by an amount
\begin{equation}\label{Eq:EHL-def}
E_{\rm HL} = \frac{2\pi^2 \gamma_s}{m d^2},
\end{equation}
which is formally negative due to the adopted convention $m<0$.

We further include the Zeeman effect resulting from an in-plane magnetic field, which we describe with the Hamiltonian
\begin{equation}\label{zeeman-def}
    H_{\rm Z} = -\kappa \left(B_+ J_- + B_-J_+\right),
\end{equation}
here, $J_{\pm} = J_x \pm iJ_y$ are the spin-3/2 raising and lowering operators, $B_\pm= B_x \pm iB_y$ where $B_x$ and $B_y$ are the components of the in-plane magnetic field, the hole $g$-factor is $\kappa$ (which is $\kappa = 3.41$ for Ge~\cite{winklerBook}), and we set $\mu_B=1$. 
Although in the main text we consider only in-plane magnetic fields, in App.~\ref{G_tensor_0D} we also incorporate the out-of-plane component $B_z$ of the field, especially relevant for the case of quantum dots addressed in Sec.~\ref{Sec:g-factor}. 
We neglect the contribution $-q{\bf B}\cdot(J_x^3,J_y^3,J_z^3)$ since for realistic values of $q$, it is typically $\sim 100$ times smaller than the contribution in Eq.~\eqref{zeeman-def}~\cite{winklerBook}. 

Finally, we add the Rashba-type SOC terms, which can arise due to asymmetries in the confining potential or to an externally applied out-of-plane electric field.
We describe this SOC by
\begin{equation}\label{RashbaSOC}
    H_{\rm R} = i\alpha_{\rm R} \left( k_+ J_- -  k_- J_+\right),
\end{equation}
where $\alpha_{\rm R}$ characterizes the strength of the coupling.

Adding the Zeeman~(\ref{zeeman-def}) and Rashba SOC terms~(\ref{RashbaSOC}) to the Luttinger--Kohn Hamiltonian projected onto the lowest tranverse momentum band, we express the full 2DHG Hamiltonian as ${\cal H}_\text{2DHG}=\sum_{\mathbf{k}}\mathcal{X}_{\mathbf{k}}^{\dagger} H_\text{2DHG}\mathcal{X}_{\mathbf{k}}$, using the $4\times 4$ matrix
\begin{widetext}
\begin{equation}\label{H2DHG}
 H_\text{2DHG}=\left(\begin{array}{cccc}
\frac{\gamma_{1}+\gamma_{s}}{2m}k^2 -\mu_h & 0 & -\sqrt{3}\left(i\alpha_{\mathrm{R}}k_{-}+\kappa B_{-}\right) & -\frac{\sqrt{3}\gamma_{s}}{2m}k_{-}^{2}\\
0 & \frac{\gamma_{1}+\gamma_{s}}{2m}k^{2}-\mu_h & -\frac{\sqrt{3}\gamma_{s}}{2m}k_{+}^{2} & \sqrt{3}\left(i\alpha_{\mathrm{R}}k_{+}-\kappa B_{+}\right)\\
\sqrt{3}\left(i\alpha_{\mathrm{R}}k_{+}-\kappa B_{+}\right) & -\frac{\sqrt{3}\gamma_{s}}{2m}k_{-}^{2} & \frac{\gamma_{1}-\gamma_{s}}{2m}k^{2}+E_{\rm HL}-\mu_h & -2\left(i\alpha_{\mathrm{R}}k_{-}+\kappa B_{-}\right)\\
-\frac{\sqrt{3}\gamma_{s}}{2m}k_{+}^{2} & -\sqrt{3}\left(i\alpha_{\mathrm{R}}k_{-}+\kappa B_{-}\right) & 2\left(i\alpha_{\mathrm{R}}k_{+}-\kappa B_{+}\right) & \frac{\gamma_{1}-\gamma_{s}}{2m}k^{2}+E_{\rm HL}-\mu_h
\end{array}\right),
\end{equation}
\end{widetext}
where $\mathcal{X}_{\mathbf{k}}=(d_{\mathbf{k},3/2}, d_{\mathbf{k},-3/2}, d_{\mathbf{k},1/2}, d_{\mathbf{k},-1/2})^{T}$ (see Fig.\ref{fig:band_structure} for illustration).
We further use the notation $k^2 = k_x^2 + k_y^2$ and $k_\pm = k_x \pm ik_y$, and we include a chemical potential $\mu_h$ for the hole gas.

\subsubsection{Superconducting film}

We consider a thin superconducting film with conventional singlet $s$-wave pairing that can be described by the BCS model.
The relevant quantum numbers for the electrons in the superconductor are the components of momentum $k_{x,y,z}$ and the spin $\sigma$.
To account for the finite thickness of the film, the momentum along the $z$-direction, $k_{z}$, is explicitly quantized, where we will use $k_{z}=\pi n_{z} / d_{{s}}$, with $n_{z}$ an integer labeling the transverse mode number and $d_{s}$ the thickness of the superconducting film.
Unlike in the semiconductor case, where we imposed a strictly two-dimensional limit (leaving only one transverse mode), the thickness of the superconductor can be arbitrary, allowing in principle all positive integers $n_{z} \geq 1$ to play a role.

Treating the attraction in the Cooper channel via a mean-field approach with the $s$-wave superconducting order parameter $\Delta$,  we arrive at a $4\times 4$ matrix Hamiltonian ${\cal H}_\text{SC}=\frac12\sum_{\mathbf{k}}\Psi_{\mathbf{k}n_z}^{\dagger}H_\text{SC}\Psi_{\mathbf{k}n_z}$, written in the Nambu basis $\Psi_{\mathbf{k}n_z}=
(c_{\mathbf{k}n_z,\uparrow},c_{\mathbf{k}n_z,\downarrow},c_{-\mathbf{k}n_z,\downarrow}^{\dagger},-c_{-\mathbf{k}n_z,\uparrow}^{\dagger})^T$, where $c_{\mathbf{k}n_z,\sigma}$
annihilates an electron with in-plane momentum $\mathbf{k} = (k_x,k_y)$, transverse mode $n_z$, and spin $\sigma$.
The matrix Hamiltonian in this basis reads explicitly as
\begin{widetext}
\begin{align}
H_\text{SC}=\left(\begin{array}{cccc}
\frac{1}{2m_s}(k^2 + \frac{\pi^2}{d_{s}^2}n_z^2) - \mu_s & -\frac{1}{2}g_sB_- & -\Delta & 0\\
-\frac{1}{2}g_sB_+ & \frac{1}{2m_s}(k^2 + \frac{\pi^2}{d_{s}^2}n_z^2) - \mu_s & 0 & -\Delta\\
-{\Delta^*} & 0 & -\frac{1}{2m_s}(k^2 + \frac{\pi^2}{d_{s}^2}n_z^2) + \mu_s & -\frac{1}{2}g_sB_-\\
0 & -{\Delta^*} & -\frac{1}{2}g_sB_+ & -\frac{1}{2m_s}(k^2 + \frac{\pi^2}{d_{s}^2}n_z^2) + \mu_s
\end{array}\right),\label{eq:hsc}
\end{align}
\end{widetext}
where $m_{s}$ is the effective mass of the electrons in the superconductor, $\mu_s$ is their
chemical potential and $g_s$ is the $g$-factor.
When writing the Hamiltonian in this form, we implicitly assume that the
Zeeman contribution from the in-plane magnetic field dominates over the orbital
contribution, as the latter is neglected. This assumption is valid for thin films, where orbital effects are suppressed.

\subsubsection{Superconductor--semiconductor coupling}

We now introduce our model of the coupling between the superconducting film and the 2DHG.
We base our approach on two natural assumptions: (1) the spin (but not the total angular momentum and its projection) of the electrons is conserved during hopping between the subsystems, which is justified if the interface is not magnetically active (e.g., does not contain magnetic impurities); and (2) the in-plane crystal momentum of the electrons is conserved during the hopping, which holds if the interface is translationally invariant on the length scale $\sim 1/k$ of the inverse typical wave vectors involved. 
Under these assumptions, the coupling Hamiltonian is
\begin{equation}
    {\cal H}_\text{hop} = \sum_{\mathbf{k},n_{z},\alpha,\sigma} 
    (t_{\alpha,n_{z}} c_{\mathbf{k}n_{z},\sigma}^{\dagger} d_{p_\alpha,\mathbf{k},\sigma} + {\rm H.c.}),
\end{equation}
where the index $\alpha \in \{x,y,z\}$ labels the $p_x$-, $p_y$-, and $p_z$-orbitals in the 2DHG, and where $\mathbf{k}$ is a two-dimensional wave vector.
In general, the hopping amplitude $t_{\alpha,n_{z}}$ depends on the orbital in the 2DHG and on the transverse mode number, $n_{z}$, in superconductor (which, in the three-dimensional limit corresponds to the dependence on the angle at which an electron in the superconductor reaches the interface). 
However, if the superconducting film is sufficiently thin, such that the transverse quantization energy $\pi^{2}/(2m_sd_{s}^{2})$ is much larger than other relevant low-energy scales (such as the order parameter $\Delta$), the leading contribution to the hopping between the superconductor and the 2DHG can come from a single \emph{resonant} transverse mode, closest in energy to the 2DHG band. In this situation contributions from other modes are suppressed by the large transverse quantization energy. In what follows we focus on such case, assuming that this resonant mode has number $n_{z}=n_{z}^{(0)}$.
The scenario where all transverse modes contribute comparably, which is relevant for thicker superconducting films, is addressed in App.~\ref{app:derive_H}. The result has the same form, and all quantitative changes can be incorporated in a renormalization of phenomenological parameters. Moreover, in experimental settings, thin superconducting layers with large transverse energy quantization are particularly relevant, as such layers are commonly used in 2D heterostructures~\cite{Vigneau2019,Valentini2024}.

Retaining only the resonant mode in the hopping term simplifies the hopping Hamiltonian,
\begin{multline}
  {\cal H}_\text{hop} = \sum_{\mathbf{k},\sigma}\big(t_{x}c_{\mathbf{k},\sigma}^{\dagger}d_{p_x,\mathbf{k},\sigma}+t_{y}c_{\mathbf{k},\sigma}^{\dagger}d_{p_y,\mathbf{k},\sigma}
  \\
  +t_{z}c_{\mathbf{k},\sigma}^{\dagger}d_{p_z,\mathbf{k},\sigma} +\text{H.c.}\big),
 \label{hop_initial}
\end{multline}
where the index $n_{z}$ is set to $n_{z}=n_{z}^{(0)}$ and omitted for
brevity.
For generality, we do not assume any specific relations between the three hopping amplitudes, except for imposing time-reversal invariance of Eq.~(\ref{hop_initial}), which guarantees that the three amplitudes $t_{x,y,z}$ can be taken to be real.

Finally, we express the hopping Hamiltonian (\ref{hop_initial}) using the appropriate angular-momentum basis states $\ket{{\bf k},J,j_z}$.
Restricting ourselves again to the $4\times 4$ subspace with total angular momentum $J=3/2$, we find the following hopping Hamiltonian:
\begin{multline}
 {\cal H}_\text{hop}=\sum_{\mathbf{k}}
 \left( -t_{+} c_{\mathbf{k}\uparrow}^{\dagger} d_{\mathbf{k},3/2} - \frac{1}{\sqrt{3}}t_{+} c_{\mathbf{k}\downarrow}^{\dagger} d_{\mathbf{k},1/2}\right. \\ 
  + t_{-} c_{\mathbf{k}\downarrow}^{\dagger} d_{\mathbf{k},-3/2} + \frac{1}{\sqrt{3}} t_{-} c_{\mathbf{k}\uparrow}^{\dagger} d_{\mathbf{k},-1/2} \\ 
    \left. +\sqrt{\frac{2}{3}} t_z c_{\mathbf{k}\uparrow}^{\dagger} d_{\mathbf{k},1/2} + \sqrt{\frac{2}{3}} t_{z} c_{\mathbf{k}\downarrow}^{\dagger} d_{\mathbf{k},-1/2} + \text{H.c.}\right),\label{eq:hopping}
\end{multline}
where $t_\pm = \frac{1}{\sqrt 2}(t_x \pm it_y)$.
We note that for a perfect lattice-matched interface the ``in-plane'' coupling amplitudes $t_\pm$ must vanish, leaving only the last two terms in Eq.~(\ref{eq:hopping}), which conserve the total in-plane angular momentum (resulting from addition of spin and orbital momenta)~\cite{Futterer2011,Moghaddam2014}.
A realistic interface will 
have lower symmetry, e.g., by being lattice-mismatched or incommensurate, and can thus allow for non-zero hopping between the $s$-orbitals and the in-plane $p_{x,y}$-orbitals of the 2DHG [see also Fig.~\ref{fig:cartoon}(a)], leading to non-zero $t_\pm$. 
In this more realistic case, the magnitude and direction of the vector ${\bf t} = (t_x,t_y,t_z)$ will strongly depend on details of the heterostructure, for which reason we will treat ${\bf t}$ as a phenomenological parameter.

\subsection{Description of the proximity effect\label{sec:3/2}}

After defining the Hamiltonians for the superconductor, the 2DHG and their coupling, in this Section we integrate  out the superconductor degrees of freedom, resulting in an effective description of the proximity effect in the 2DHG. 
In the first step, we derive the self-energy for the electrons in the semiconductor, from which we obtain an effective low-energy Hamiltonian for the 2DHG in the heavy-hole and light-hole subspace, which is the main general result of this work. 
In the second step, we project this Hamiltonian using a Schrieffer--Wolff transformation onto the heavy-hole subspace~\cite{lidal23}, which is the relevant subspace for many low-density quantum-technological applications~\cite{ScappucciReview}.

\subsubsection{Self-energy for the 2DHG}

In this subsection, we derive the self-energy for the 2DHG that arises due to coupling to the superconducting film via the hopping Hamiltonian (\ref{eq:hopping}). This calculation generalizes the self-energy expression  used for a single-band semiconductor in tunneling contact with a superconductor \cite{Stanescu2010,Sau2010}. We start from the total partition function of the coupled system, written as a functional path integral~\cite{Altland_Simons_2023},
\begin{multline}
 Z=\intop{\cal D}\left[\psi,\overline{\psi},\chi,\overline{\chi}\right]
 \exp\Big\{
 \sum_{n} \Big(i\epsilon_{n}(\overline{\psi}_{n}\psi_{n}+\overline{\chi}_{n}\chi_{n})
 \\
 -{\cal H}_\text{SC}\left[\overline{\psi},\psi\right]-{\cal H}_\text{2DHG}\left[\overline{\chi},\chi\right]-{\cal H}_\text{hop}\left[\overline{\psi},\psi,\overline{\chi},\chi\right]\Big)
\Big\},
\label{partition_all}
\end{multline}
where we introduced the Grassmann fields $\psi,\overline{\psi}$ and $\chi,\overline{\chi}$ which correspond to operators $c,c^{\dagger}$ and $d,d^{\dagger}$ respectively.
Note that in the term $\sum_n i\epsilon_{n}(\overline{\psi}_{n}\psi_{n}+\overline{\chi}_{n}\chi_{n})$, we only explicitly show the summation over Matsubara frequencies, $\epsilon_{n}=2\pi\left(n+\frac{1}{2}\right)$, while suppressing the summations over the indices $\{ \mathbf{k},j_z\}$ for the semiconductor fields $\chi$ and over the indices $\{\mathbf{k},n_{z},\sigma\}$ for the superconductor fields $\psi$; additionally, we omit the index $n$ in the last three terms for brevity.
We also note that the term ${\cal H}_\text{SC}\left[\overline{\psi},\psi\right]$ includes all transverse modes $n_{z}$ in superconductor, while the term ${\cal H}_\text{hop} \left[ \overline{\psi}, \psi,\overline{\chi}, \chi\right]$ contains only the resonant mode, as discussed in the previous section.

By integrating out the superconducting degrees of freedom, which is equivalent to calculating the Gaussian integral over $\psi,\overline{\psi}$ in Eq.~(\ref{partition_all}) (see App.~\ref{app:derive_H}) we derive
\begin{multline}
    Z=\intop{\cal D}\left[\chi,\overline{\chi}\right]
    \exp\Big\{
    \sum_{n}\Big(\sum_{\mathbf{k}j}i\epsilon_{n} \overline{\chi}_{\mathbf{k}nj}\chi_{\mathbf{k}nj} \\
    -
    {\cal H}_\text{2DHG}[\overline{\chi},\chi]-\frac{1}{2}
\sum_{\mathbf{k}} \overline{\Theta}_{\mathbf{k}n} G_{\rm SC}(i\epsilon_n,{\bf k})
\Theta_{\mathbf{k}n}
 \Big)
    \Big\},
\label{partition_2DHG}
\end{multline}
where the following notations are used: $G_{\rm SC}(i\epsilon_n,{\bf k})$~is the Green function of the superconductor,
\begin{equation}
    G_{\rm SC}^{-1}(i\epsilon_n,{\bf k}) = i\epsilon_{n}-H_\text{SC}|_{n_z = n_z^{(0)}},
\end{equation}
with $H_\text{SC}$ as defined in Eq.~(\ref{eq:hsc}).
The field $\Theta_{\mathbf{k}n}$ emerges due to the hopping between the superconductor and the 2DHG [Eq.~(\ref{eq:hopping})] and is defined as
\begin{equation}
    \Theta_{\mathbf{k}n}=\big( 
\theta_{\mathbf{k}n,1},-\theta_{\mathbf{k}n,2}, \overline{\theta}_{-\mathbf{k}-n,2},\overline{\theta}_{-\mathbf{k}-n,1}\big)^T,
\end{equation}
where $\theta_{\mathbf{k}n,1}=-t_{+}\chi_{\mathbf{k}n,3/2}+\frac{1}{\sqrt{3}}t_{-}\chi_{\mathbf{k}n,-1/2}+\sqrt{\frac{2}{3}}t_{z}\chi_{\mathbf{k}n,1/2}$
and $\theta_{\mathbf{k}n,2}=\frac{1}{\sqrt{3}}t_{+}\chi_{\mathbf{k}n,1/2}-t_{-}\chi_{\mathbf{k}n,-3/2}-\sqrt{\frac{2}{3}}t_{z}\chi_{\mathbf{k}n,-1/2}$. The term $\frac{1}{2}\sum_{\mathbf{k}}\overline{\Theta}_{\mathbf{k}n}G_{SC}\Theta_{\mathbf{k}n}$ in Eq.~(\ref{partition_2DHG}) represents the self-energy correction to the Hamiltonian, encoding all relevant information about the effect of the superconducting film on the 2DHG, including the proximity effect.

In what follows, we introduce doubled bispinors to account for the doubling of the spinor space resulting from the use of Nambu space,
\begin{subequations}
    \label{bibispinor}
\begin{align}
\check{{\cal X}}_{\mathbf{k}n}
&=\left(
{\cal X}_{\mathbf{k}n}^{(3/2)},
{\cal X}_{\mathbf{k}n}^{(1/2)}
\right)^T,
\\
{\cal X}_{\mathbf{k}n}^{(j)}&=\left(\chi_{\mathbf{k}n,j},\chi_{\mathbf{k}n,-j},\overline{\chi}_{-\mathbf{k},-n,-j},-\overline{\chi}_{-\mathbf{k},-n,j}\right)^T,
\end{align}
\end{subequations}
where $j=3/2$ and $1/2$ labels the HH and LH bands. 
We emphasize that here we group together spinors corresponding to the HH and LH bands separately, since such ordering facilitates the projection onto the HH band implemented below. 
In contrast, Sec.~\ref{sec:summary} used a more conventional basis, grouping together all particle and anti-particle operators, see Eq.~(\ref{time_rev_basis}).

We can now express the self-energy correction as
\begin{equation}
\sum_{\mathbf{k}}\overline{\Theta}_{\mathbf{k}n}G_{SC}\Theta_{\mathbf{k}n}=\sum_{\mathbf{k}}\overline{\check{\cal X}}_{\mathbf{k}n}\Sigma \check{{\cal X}}_{\mathbf{k}n},
\label{self-energ_Sigma}
\end{equation}
where the $8\times8$ matrix $\Sigma$ is the self-energy in the basis~(\ref{bibispinor}) (see App.~\ref{app:derive_H} for an explicit expression). Particularly, retaining in $\Sigma$ only the elements corresponding to the induced superconducting pairing, i.e., proportional to $\Delta$, (denoted as $\Sigma_{\Delta}$), we derive,
\begin{equation}\label{sigma}
  \Sigma_{\Delta}=\left(\begin{array}{cccc}
0 & -{\Delta}_{H}\mathbb{I}_{2\times2} & 0 & \hat{\Delta}_{HL}\\
-{\Delta}_{H}^{\dagger} \mathbb{I}_{2\times2} & 0 & \sigma_{y}\hat{\Delta}_{HL}^{*}\sigma_{y} & 0\\
 0 & \sigma_{y}\hat{\Delta}_{HL}^{T}\sigma_{y} & 0 & {\Delta}_{L}\mathbb{I}_{2\times2}\\
\hat{\Delta}_{HL}^{\dagger} & 0 & {\Delta}_{L}^{\dagger}\mathbb{I}_{2\times2} & 0
\end{array}\right),
\end{equation}
where $\mathbb{I}_{2\times2}$ is the $2\times2$ identity matrix, and the matrix $\hat{\Delta}_{HL}$ contains two additional pairing parameters inside,
\begin{equation}
    \hat{\Delta}_{HL}
    =
    \left(\begin{array}{cc}
 \Delta_{HL,a} & \Delta_{HL,b} \\
 \Delta_{HL,b}^*&-\Delta_{HL,a}^*
 \end{array}\right).
\end{equation}

Our derivations in the Appendix give explicit expressions for all four pairings appearing in Eq.~(\ref{sigma}),
\begin{subequations}\label{Eq:pairings-all}
\begin{eqnarray}
 {\Delta}_{H} & = & -\Delta \tau_{+-},\\
 {\Delta}_{L} & = & -\frac{\Delta}{3}(\tau_{+-} + 2 \tau_{zz} ),\\
 {\Delta}_{HL,a} & = & \sqrt \frac{2}{3}\Delta  \tau_{-z},\\
 {\Delta}_{HL,b} & = & \frac{\Delta}{\sqrt 3} \tau_{--},
\end{eqnarray}
\end{subequations}
where the dimensionless parameters $\tau_{\alpha \beta}$
\begin{equation} \label{hop_strength}
    \tau_{\alpha \beta} = \frac{t_\alpha t_\beta }{\left|\Delta\right|^{2}+\xi_{k,s}^{2}-\varepsilon^{2}},
\end{equation}
characterize the strength of the proximity effect.
Although $\tau_{\alpha\beta}$ is a tensor with 9 components, it is fully specified by the choice of the three hopping parameters in the vector $\mathbf{t} = (t_x,t_y,t_z)$.
Note that we have moved from Matsubara frequencies $i\epsilon_n$ to real energies $\varepsilon$ and we introduced $\xi_{k, s} = k^2/2m_s + \pi^2 (n_z^{(0)})^2/(2m_s d^2_s) - \mu_s$ which is the $k$-dependent energy of the electrons in the resonant transverse mode of the superconductor, measured from the Fermi level. 

The expressions in Eq.~(\ref{Eq:pairings-all}) provide explicit values for the induced pairings in terms of hopping and other microscopic parameters, and constitute the main result of our work. 
In particular, upon a permutation of the basis from Eq.~(\ref{bibispinor}) to the basis in Eq.~(\ref{time_rev_basis}), the self-energy matrix~(\ref{sigma}) obtained in this section takes precisely the form of the pairings introduced in Sec.~\ref{sec:summary}, see Eq.~(\ref{pairLHHHtrans}).
Physically, the results obtained from more phenomenological considerations in Sec.~\ref{sec:summary} and derived more thoroughly here imply that the induced superconductivity can result in both intraband contributions $ {\Delta}_{H}$, ${\Delta}_{L}$, and interband contributions ${\Delta}_{HL,a}$, ${\Delta}_{HL,b}$.
In the limit of $t_\pm = 0$, only the LH intraband pairing term ${\Delta}_{L}$ remains, which again reflects that in this case the total in-plane angular momentum has to be conserved during tunneling between the two subsystems.

\subsubsection{Effective pairing Hamiltonian for heavy holes}
\label{sec:effham}

We now derive and analyze the effective model for the HH $j_z = \pm 3/2$ subspace spanned by the states $\ket{{\bf k},\pm \frac{3}{2}}$.
The partition function in Eq.~(\ref{partition_2DHG})
provides all relevant information about the proximitized 2DHG, enabling the explicit derivation of the effective Green function.
Since we introduce Nambu space to account for the induced pairing, the self-energy $\Sigma$, defined by Eq.~(\ref{self-energ_Sigma}), is an $8\times8$ matrix which, in turn, results in an $8\times8$ Green function matrix (see the details in App.~\ref{app:derive_H}).

While this matrix can be studied further to understand the total proximity effect in the combined HH and LH bands, we instead proceed to simplify the model, by assuming a large HH--LH splitting and focusing on the dynamics of the HHs only.
Indeed, in typical experiments the confinement along $\hat z$ is strong enough to split the two bands by $E_{\rm HL} \sim 20$~meV, yielding a two-dimensional low-energy HH subspace where the chemical potential crosses only the HH band, and the corresponding Fermi energy is much smaller than the band splitting $E_{\rm HL}$. 
We thus focus on the HH $j_z = \pm 3/2$ band and take into account the effect of the coupling between the HH and LH bands perturbatively.

To derive an effective model for the HH band, we apply a Schrieffer--Wolff
transformation to the Green function, projecting it onto the $\left|{\bf k}, \pm \frac{3}{2}\right\rangle$ subspace, see App.~\ref{app:derive_H} for details.
This results in an effective $4\times4$ Green function $G$ in HH Nambu space of the form
\begin{align}
G^{-1}=\left(1+\tau_{+-} \right)\varepsilon-H,
\label{Green_func_3/2}
\end{align}
where $\varepsilon$ is a real energy.
The $\varepsilon$-dependence of the matrix $H$ is weak compared to that of the leading contribution $(1+\tau_{+-})\varepsilon$, and can therefore be neglected. 
Consequently, $H/(1+\tau_{+-} )$ can be treated as an effective Hamiltonian. 
For simplicity, we do not consider the factor $1/(1+\tau_{+-} )$ in our analysis below, but do take it into account when calculating observables in the next section.

The explicit matrix form of the effective HH Hamiltonian in Eq.~(\ref{Green_func_3/2}) is given by
\begin{align}
\label{eff_Hamilton}
     & H=\left(\begin{array}{cc}
\hat{H}\left(\mathbf{k}\right) & \hat{\Delta}\left(\mathbf{k}\right)\\
\hat{\Delta}^{\dagger}\left(\mathbf{k}\right) & -\sigma_{y}\hat{H}\left(-\mathbf{k}\right)^{T}\sigma_{y}
\end{array}\right),
\end{align}
where $\hat{H}\left(\mathbf{k}\right)$ and $\hat{\Delta}\left(\mathbf{k}\right)$ are $2\times2$ matrices. In order to simplify expressions for the individual components of this Hamiltonian listed below, we introduce three dimensionless parameters  
\begin{eqnarray} \label{zetas}
 \zeta_{\rm F}=\frac{\gamma_s k^2}{mE_{\rm HL}},\quad
 \zeta_{\rm R}=\frac{2\alpha_{\rm R}k}{E_{\rm HL}},\quad
 \zeta_B=\frac{2\kappa B}{E_{\rm HL}},
\end{eqnarray}
that correspond to the characteristic energy scales of HH--LH hole mixing terms due to kinetic energy, Rashba coupling, and Zeeman coupling compared to the energy scale of the HH--LH splitting. 
We note that although $\zeta_{\rm F,R}$ are still functions of $k$, we will often restrict the momentum to the Fermi surfaces, making them effectively constants.

\begin{figure}[tb]
\includegraphics[width=\linewidth]{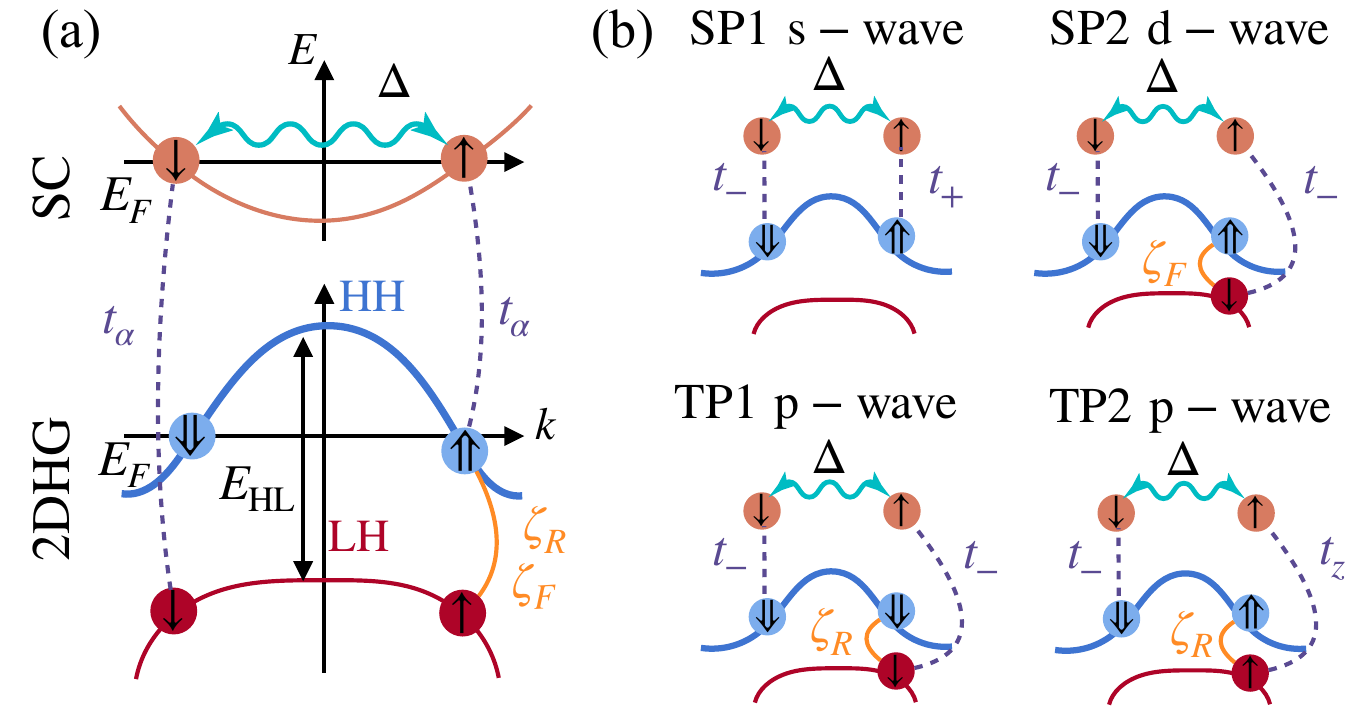}
\caption{(a) Structure of the cartoons we use to represent the processes underlying the different proximity-induced corrections to the 2DHG Hamiltonian.
The band structures of the superconductor and the 2DHG are sketched in orange (SC), blue (HHs), and red (LHs). 
The blue wavy line in the superconducting band at the top is the $s$-wave spin-singlet pairing in the superconductor.
The dashed lines marked $t_\alpha$ with $\alpha=\pm, z$ indicate hopping processes between the superconductor and the 2DHG.
The orange lines indicate processes that  mix HH and LH states, such as the off-diagonal terms of the kinetic energy ($\zeta_{\rm F}$) or Rashba SOC ($\zeta_{\rm R}$).
We focus on the proximity effect in the HH band, where the states with angular momentum projection $j_z = \pm3/2$ are denoted by double arrows. 
The LH band, where the states have $j_z = \pm 1/2$ (single arrows), is split off by the energy $E_\text{HL}$.
(b)~Processes underlying the four terms in Eqs.~(\ref{Eq:D11},\ref{Eq:D12}). 
The $s$-wave pairing term SP1 results from the correlated tunneling of a Cooper pair into HH band.
The $d$-wave singlet pairing term SP2 originates from the two particles of a Cooper pair tunneling into a HH and a LH state, where the LH state has a finite HH character due to the band mixing quantified by $\zeta_{\rm F}$.
Finally, the two triplet pairings TP1 and TP2 result from similar processes, but relying on band mixing by Rashba SOC.}\label{fig:diagrams} 
\end{figure}

Using the notations introduced in Eq.~(\ref{zetas}), we can write the four components of the pairing matrix $\hat\Delta({\bf k})$ in the effective HH Hamiltonian~(\ref{eff_Hamilton}) as
\begin{align}
 \label{Eq:D11}
\Delta_{11}\left(\mathbf{k}\right)
= {} & {} 
 \underbrace{\Delta\tau_{+-} + i\sqrt{2}\Delta|\tau_{z+}|\zeta_B\sin\left(\phi_{B}-\phi_{t}\right)}_{\text{SP}1}
 \nonumber \\
 {} & {} \underbrace{{} 
 -{}\Delta\tau_{+-}\zeta_F\cos\left[2\left(\phi_{\mathbf{k}}-\phi_{t}\right)\right]}_{\text{SP}2} \nonumber\\
 {} & {} \underbrace{{} - {} \sqrt{2}\Delta|\tau_{z+}| \zeta_R\sin\left(\phi_{\mathbf{k}}-\phi_{t}\right)}_{\text{TP}2},
 \\
 \Delta_{12}\left(\mathbf{k}\right)= {} & {} 
 \underbrace{{}-{}i\Delta\tau_{+-} \zeta_R e^{-i\left(\phi_{\mathbf{k}}+2\phi_{t}\right)}}_{\text{TP}1}, \label{Eq:D12} 
\end{align}
where $\phi_{\mathbf{k}}$ is the in-plane angle of the momentum $\mathbf{k}$, $\phi_{t}$ denotes the phase of the hopping constant $t_{+} = \left|t_{+}\right|e^{i\phi_{t}}$, and $\phi_{B}$ is the in-plane angle of the magnetic field.
We do not impose any specific restrictions on the parameters $\tau_{\alpha\beta}$ and $\phi_t$, which characterize the hopping, to maintain the generality of the model. 
The remaining two components follow as $\Delta_{22}(\mathbf{k}) = \Delta_{11}(-\mathbf{k})$ and $\Delta_{21}(\mathbf{k}) =\Delta_{12}^{*}(\mathbf{k})$.

Next, we analyze the physical meaning of these pairing terms.  
The simplest term is  the $s$-wave singlet pairing contribution (SP1), given by the first term on the first line of Eq.~(\ref{Eq:D11}) (we do not discuss the magnetic-field-induced correction represented by the second term in SP1).
This contribution simply arises from direct tunneling of Cooper pairs from superconductor into the HH band, allowed by finite $t_\pm$, see the corresponding cartoon in Fig.~\ref{fig:diagrams}(b). 
Next, the second line of Eq.~(\ref{Eq:D11}) contains induced $d$-wave singlet pairing denoted as SP2. This term, also illustrated in Fig.~\ref{fig:diagrams}(b), requires the more intricate process when one of the carriers of the Cooper pair tunnels into the HH band $|j_z|=3/2$ and the other into the LH $|j_z|=1/2$ band, after which it gets admixed into the HH band through interband elements of the kinetic energy which results in the proportionality of the term to $\zeta_{\rm F}$.
This term has a nontrivial dependence on the angle $\phi_{\bf k}$ along the Fermi surfaces, thus leading to a qualitative change of the density of states, as we will explore in Sec.~\ref{sec:observables} below.
Finally, the last line of Eq.~(\ref{Eq:D11}) and Eq.~(\ref{Eq:D12}) contain two different induced triplet order parameters denoted as TP1 ($p_{x}\pm ip_{y}$-type) and TP2 ($p_z$-type), and they are illustrated on the  bottom row of Fig.~\ref{fig:diagrams}(b).
These processes also rely on Cooper pairs splitting up into a HH and a LH state, where in this case the admixture of the LH state into the HH band is facilitated by Rashba SOC, explaining the proportionality of these terms to the parameter $\zeta_{\rm R}$.

The proximity to the superconductor also yields corrections to the diagonal blocks $\hat H({\bf k})$ in the HH Hamiltonian, describing a renormalization of the SOC and the Zeeman effect.
These renormalizations are described by the effective Hamiltonian~(\ref{eff_Hamilton}), where
\begin{widetext}
\begin{align}
 & H_{11}\left(\mathbf{k}\right)=H_{22}\left(-\mathbf{k}\right)\Big|_{\mathbf{B}\rightarrow-\mathbf{B}}=\xi_{{k}}+\underbrace{\xi_{k_{\rm F},s}\tau_{+-}\zeta_{\rm F} \cos\left[2\left(\phi_{\mathbf{k}}-\phi_{t}\right)\right]}_{\text{FSA}}\underbrace{{} -{} \left(3\kappa 
 + \tfrac{1}{2}g_s \tau_{+-} \right) B \zeta_{\rm R}\sin\left(\phi_{\mathbf{k}}-\phi_{B}\right)}_{\text{FSS}}\nonumber \\
 & \hspace{4em}
 + \underbrace{\sqrt{2}\alpha_{\rm R} k |\tau_{z+}|\zeta_{s}  \sin\left(\phi_{\mathbf{k}}-\phi_{t}\right)}_{\text{R}3}
 + \underbrace{\sqrt{2} \kappa|\tau_{z+}| B\zeta_{s}  \cos\left(\phi_{B}-\phi_{t}\right)}_{\text{Z}3}
 + \underbrace{\tfrac{1}{2}\sqrt{2} g_s|\tau_{z+}| B  \zeta_{\rm F} \cos\left(2\phi_{\mathbf{k}}-\phi_{t}-\phi_{B}\right)}_{\text{Z}4},\label{H11}\\
 & H_{12}\left(\mathbf{k}\right)=\underbrace{{}-{}3i\alpha_{\rm R}k \zeta_{\rm F} e^{-3i\phi_{\mathbf{k}}}}_{\text{R}1} + \underbrace{i\alpha_{\rm R} k \tau_{+-} \zeta_{s}  e^{-i\left(\phi_{\mathbf{k}}+2\phi_{t}\right)}}_{\text{R}2}\underbrace{{}-{}\left(3\kappa +\tfrac{1}{2}g_s \tau_{+-} \right) B \zeta_{\rm F} e^{-i(2\phi_{\mathbf{k}}+\phi_{B})}}_{\text{Z}1} + \underbrace{\tfrac{1}{2}g_s\tau_{+-} B e^{-i\left(\phi_{B}+2\phi_{t}\right)}}_{\text{Z}2}.
 \label{H12}
\end{align}
\end{widetext}
Here we introduced the 2DHG kinetic energy $\xi_{{k}}=k^{2}\left(\gamma_{1}+\gamma_{s}\right)/2m-\mu_h$ and the dimensionless parameter
\begin{equation}
    \zeta_s = \frac{2 \xi_{k_{\rm F},s} }{E_{\rm HL}},
\end{equation}
where $\xi_{k_{\rm F},s}=\xi_{{k},s}|_{k=k_{{\rm F},\text{2DHG}}}$ is the kinetic energy of electrons in the superconductor with the Fermi wave vector $k_{\rm F}$ of the 2DHG.

\begin{figure}[t]
\includegraphics[width=0.85\linewidth]{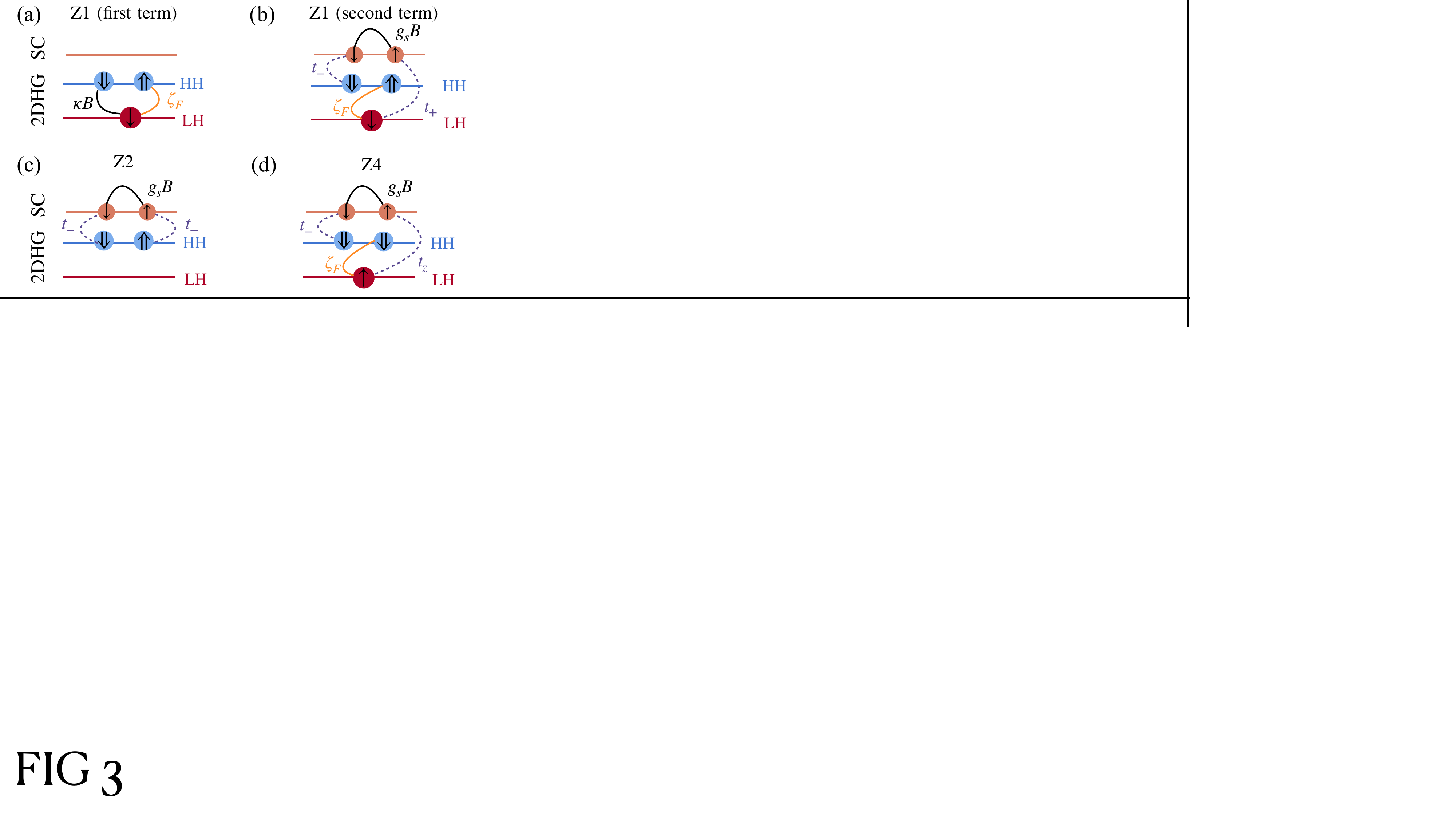}
\caption{\label{fig:diagrams_Z} 
Illustration of processes that induce corresponding Zeeman terms in Eqs.~(\ref{H12})-(\ref{H11}). We use the same notations as in Fig.~\ref{fig:diagrams}, however do not show dispersion and introduce solid black line to show Zeeman field in 2DHG ($\kappa B$) and superconductor ($g_sB$). Panels (a)-(b) show two contributions to Z1 term: (a) intrinsic Zeeman coupling of 2DHG, where magnetic field changes spin (b) Zeeman coupling induced by tunneling into superconductor and band mixing, $\zeta_F$. (c)-(d) Show the terms Z2 in Eq.~(\ref{H12}) and Z4 in Eq.~(\ref{H11}), in both cases Zeeman field acts within superconductor, and coupling is induced either by direct hopping processes into heavy hole band as in (c), or (d) by hopping from heavy hole band into the superconductor, and then into light hole band, using band mixing process.}
\end{figure}

Similar as before, we can analyze the physical meaning of the contributions representing the effective Rashba SOC (R1--R3), the Zeeman effect (Z1--Z4), and the deformation of Fermi surfaces (FSA, FSS).
The term R1 in $H_{12}$ corresponds to the conventional Rashba SOC for the HH band which is cubic in momentum (and thus $\propto e^{-3i\phi_\mathbf{k}}$)~\cite{Fang2023} and is not affected by the proximity of the superconductor.
The terms R2 and R3 represent linear-in-momentum contributions ($\propto e^{\pm i\phi_{\mathbf{k}}}$) to Rashba SOC that arise due to the coupling with the superconductor. Specifically, these terms emerge when a HH tunnels into the superconductor, returns as a LH, and subsequently mixes into the HH band again through the Rashba SOC.
As a result, a dependence on the kinetic energy of electrons in the superconductor, $\xi_{k_{\rm F},s}$, and proportionality to the hopping strengths $\tau_{+-}$ and $|\tau_{z+}|$ are acquired during this process.

Next, we consider the Zeeman terms (Z1--Z4). 
The first term in Z1 (proportional to $\kappa$) represents the conventional in-plane Zeeman effect for the HH band, which is independent of the coupling to the {superconductor}. 
In Fig.~\ref{fig:diagrams_Z}(a) we illustrate how this term arises as a second-order process involving interband Zeeman coupling and HH--LH mixing by interband contributions to the kinetic energy.
The second term in Z1 arises through the following process, illustrated in Fig.~\ref{fig:diagrams_Z}(b): a HH tunnels into the superconductor, experiences a Zeeman effect $\propto g_s B$, tunnels back into the 2DHG as a LH, and then gets admixed into the HH band via HH--LH mixing induced by interband elements of the kinetic energy.
The term Z2, illustrated in Fig.~\ref{fig:diagrams_Z}(c), arises from the process of a HH tunneling into the superconductor, experiencing a Zeeman effect $\propto g_s B$, and tunneling back into the 2DHG as a HH again.
We note that the direction of the effective Zeeman field for this term is rotated in-plane by an angle $2\phi_{t}$, where $\phi_{t}$ is the phase of the hopping parameter $t_+ = \frac{1}{\sqrt 2}(t_x + it_y)$ [see also the paragraph below Eq.~(\ref{Eq:D12})],
due to the tunneling-induced change in angular momentum associated with the tunnel coupling parameters $t_{\pm,z}$. 
Finally, the last two induced Zeeman terms, Z3 and Z4, appear in the diagonal components of the Hamiltonian and represent the component along $z$ of the effective magnetic field.
The term Z3 arises from a process similar to R3 (not illustrated for brevity), whereas the term Z4 results from the process illustrated in Fig.~\ref{fig:diagrams_Z}(d), which is very similar to Z1 but with the LH tunneling back into the 2DHG (and thus also the final HH) having opposite spin.
This yields a contribution to an effective Zeeman field along the $z$-axis as well. 
 
To conclude the discussion of the effective Hamiltonian, we examine the remaining components in $H_{11}$ labeled FSA and FSS. 
The first term (FSA) leads to a renormalization of the effective mass tensor, resulting in a Fermi surface anisotropy. 
The second term (FSS) corresponds to a Fermi surface shift, induced by the magnetic field. 
This shift is perpendicular to the magnetic field direction, and can be attributed to the simultaneous presence of SOC and a magnetic field. 

\section{From effective Hamiltonian to physical properties\label{sec:observables}}

In this section we illustrate some of the possible applications of the effective Hamiltonian derived above. 
To this end, we calculate several physical quantities, highlighting observable consequences of the proximity effect as described above, in both two-dimensional and zero-dimensional (quantum dot) settings. 
First, in Sec.~\ref{sec:Rashba_split}, we further simplify the Hamiltonian assuming that strong Rashba SOC results in a large enough band splitting such that interband couplings can be neglected.
Then, in Sec.~\ref{Bb_DOS}, we analyze the Bogoliubov bands, the density of states (DOS), and the Bogoliubov Fermi surfaces (BFSs) that appear for sufficiently strong magnetic fields. 
Finally, in Sec.~\ref{Sec:g-factor} we calculate corrections to the effective $g$-tensor in a quasi-zero-dimensional geometry and predict its dependence on material parameters and the geometry of the quantum dot.

\subsection{Simplified heavy-hole Hamiltonian\\ for strong Rashba SOC}
\label{sec:Rashba_split}

Here we assume strong Rashba SOC, such that the resulting Rashba-split bands are sufficiently separated to allow to approximate the $4 \times 4$ Hamiltonian~(\ref{eff_Hamilton}) as
\begin{align}
\label{H_simp}
 & \tilde{H}=\left(\begin{array}{cc}
H^{(-)} & 0\\
0 & H^{(+)}
\end{array}\right),
\end{align} 
when transformed to a suitable basis (see for the details App.~\ref{app:Rashba_split}).
The two $2\times 2$ Hamiltonians operate in the Nambu spaces of the two separate Rashba-split bands and read explicitly as
\begin{equation}\label{Eq:HRashba}
H^{(\pm)}=\left(\begin{array}{cc}
\Xi_{\mathbf{k}}^{(\pm)}\mp V_{\mathbf{k}}^{(\pm)} & \Delta_{\mathbf{k}}^{(\pm)}\\
\vphantom{\Bigg(}\left(\Delta_{\mathbf{k}}^{(\pm)}\right)^* & -\Xi_{\mathbf{k}}^{(\pm)}\mp V_{\mathbf{k}}^{(\pm)}
\end{array}\right),
\end{equation}
where the band dispersion $\Xi_{\mathbf{k}}^{(\pm)}$, the Zeeman terms $V_{\mathbf{k}}^{(\pm)}$, and the effective pairing $\Delta_{\mathbf{k}}^{(\pm)}$ are
\begin{align}
\Xi_{{k}}^{(\pm)} &= 
\xi_{\mathbf{k}} \mp 3\alpha_{\rm R}k\zeta_{\rm F} \nonumber\\ \label{Eq:xi2x2}
& +\xi_{k_{\rm F},s}\tau_{+-} \left(\zeta_{\rm F}\pm\zeta_{\rm R}\right)\cos [2(\phi_{\mathbf{k}}-\phi_{t})],\\
\nonumber
V_{\mathbf{k}}^{(\pm)}&=\frac{1}{2}g_s B \tau_{+-} \sin\left(\phi_{B}-3\phi_{\mathbf{k}}+2\phi_{t}\right) \\
\label{Eq:Zeeman2x2}
& -\frac{1}{2}\left(6\kappa + g_s \tau_{+-}\right)B\left(\zeta_{\rm F}\pm\zeta_{\rm R}\right)\sin\left(\phi_{B}-\phi_{\mathbf{k}}\right),\\
\label{Eq:Delta2x2}
 \Delta_{\mathbf{k}}^{(\pm)}&=\Delta\tau_{+-}\left[1 - \left(\zeta_{\rm F}\pm\zeta_{\rm R}\right)\cos\left(2\phi_{\mathbf{k}}-2\phi_{t}\right)\right],
\end{align}
see App.~\ref{app:Rashba_split} for a derivation.
The most important terms in Eq.~(\ref{Eq:xi2x2}) are the bare HH dispersion $\xi_{\mathbf{k}}$ and the (cubic) Rashba energy splitting of $\pm 3\alpha_{\rm R}k\zeta_{F}$. 
The Zeeman term~(\ref{Eq:Zeeman2x2}) has two types of contributions, one coming from the intrinsic Zeeman effect in the 2DHG (proportional to $\kappa$), which emerges due to the HH--LH mixing, and one from the coupling to the superconductor (proportional to $g_s$).
Finally, the effective pairing~(\ref{Eq:Delta2x2}) has a constant as well as momentum-dependent contributions, the latter originating from the $d$-wave singlet and the triplet pairings in the effective Hamiltonian (\ref{eff_Hamilton}). 
Notably, the contribution from the triplet pairing makes the momentum-dependent pairing term different in the two Rashba-split bands.

\begin{figure*}[t]
\includegraphics[width=\linewidth]{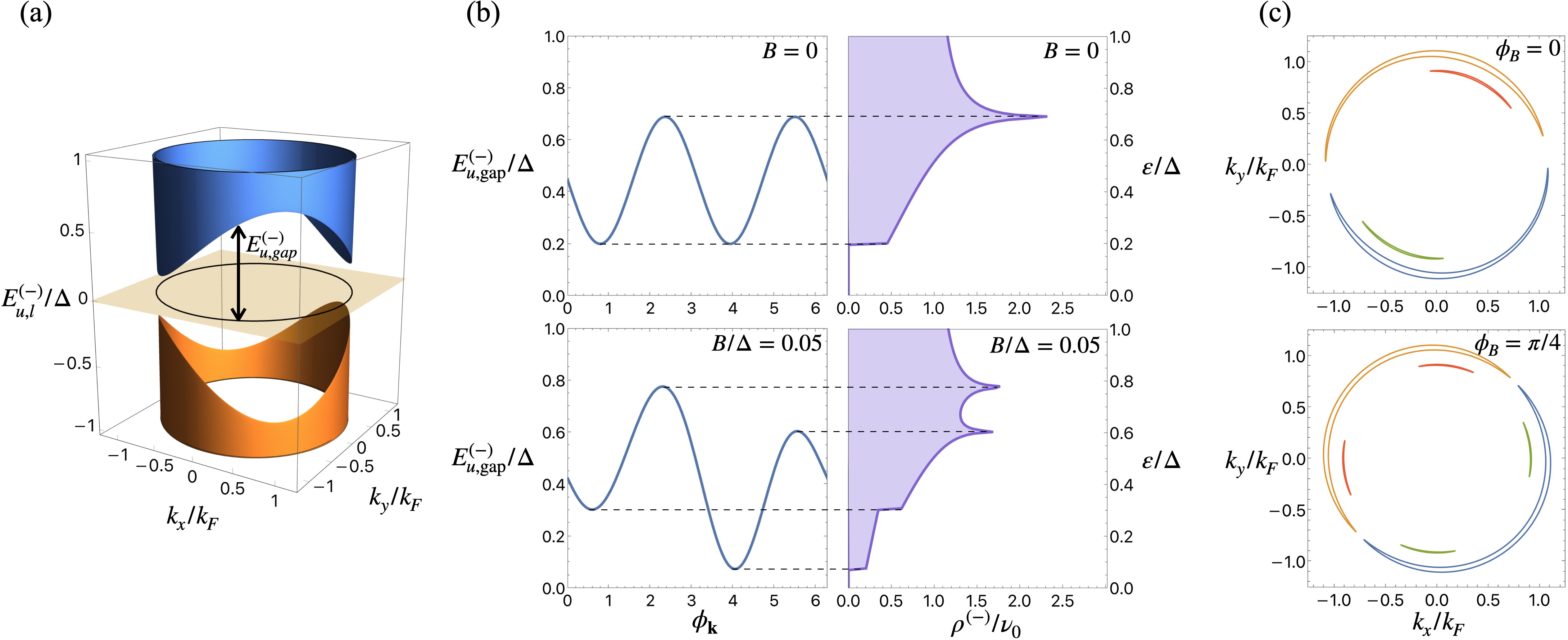}
\caption{\label{Fig:obs}
(a) Bogoliubov bands $E^{(-)}_{u,l}$ as a function of two-dimensional momentum ${\bf k}$ for $B=0$, as calculated from Eq.~(\ref{energy_spectrum}), \edits{illustrating the effect of the induced superconductivity on the one of the two spin--orbit-split bands}.
The bands feature a $\phi_{\bf k}$-dependent distance between the band extrema and the Fermi level ($\phi_{\mathbf{k}}$ is the in-plane angle of the momentum $\mathbf{k}$), \edits{reflecting the unconventional nature of the superconductivity}.
(b) \edits{Illustration of the effect of an additional Zeeman field on the bands.} Left: The distance of the bottom of the upper band shown in (a) to the Fermi level, $E^{(-)}_{u,\text{gap}}$, as a function of $\phi_{\bf k}$, for two values of in-plane magnetic field $B=0$ and $0.05\,\Delta$ with $\phi_B = 0$ ($\phi_{B}$ is the in-plane angle of the magnetic field).
Right: The DOS of the upper band has singularities at energies where $E^{(-)}_{u,\text{gap}}$ has extrema.
In contrast to square-root BCS singularities, we observe discontinuities in the DOS at small energies and logarithmic peaks at larger energies. 
The location and number of singularities in the DOS are tunable by the in-plane magnetic field. 
(c) For even stronger magnetic field ($B=0.7\,\Delta$), the gap in the \emph{total} DOS $\rho_+(\varepsilon) + \rho_-(\varepsilon)$ closes and Bogoliubov Fermi surfaces emerge.
\edits{The two concentric ring-like features correspond to the contributions from the two spin--orbit split bands.}
From comparing two different orientations of magnetic field: along the $x$-axis ($\phi_B = 0$, top) and along $x$-$y$ diagonal ($\phi_B=\pi/4$, bottom) we see that the shape of the surfaces depends sensitively on the direction of the magnetic field.
}
\end{figure*}

The energy spectrum of the Hamiltonian~(\ref{H_simp}) can be calculated explicitly, yielding
\begin{equation}
E_{u}^{(\pm)}({\bf k}) =  \frac{\sqrt{(\Xi_{\mathbf{k}}^{(\pm)})^{2}+|\Delta_{\mathbf{k}}^{(\pm)}|^{2}}\mp V_{\mathbf{k}}^{(\pm)}}{1+\tau_{+-}},
\label{energy_spectrum}
\end{equation}
for the two upper Bologiubov bands, where we included the renormalization factor $1/(1+\tau_{+-})$.
The lower bands $E_{l}^{(\pm)}({\bf k})$ are obtained from (\ref{energy_spectrum}) by inverting only the sign of the square root. 
Finally, the simple form of the projected Hamiltonian allows to calculate the density of states per spin for the two Rashba-split bands,
\begin{equation}
     \label{DOS}
\rho_{\pm}\left(\varepsilon\right)=\nu_{0}\intop\frac{d\phi_{\mathbf{k}}}{2\pi}
|\mho^{(\pm)}_{\mathbf{k}}|\frac{\theta[(\mho_{\mathbf{k}}^{(\pm)})^{2}-|\Delta_{\mathbf{k}}^{(\pm)}|^{2}]}{\sqrt{\left|(\mho_{\mathbf{k}}^{(\pm)})^{2}-|\Delta_{\mathbf{k}}^{(\pm)}|^{2}\right|}},
\end{equation}
where $\nu_0$ is the density of states of the 2DHG at the Fermi level for one projection of angular momentum, $\mho^{(\pm)}_{\mathbf{k}} = (1+\tau_{+-})\varepsilon \pm V_{\mathbf{k}}^{(\pm)}$, 
$\theta(x)$ is the Heaviside step function, 
and $V_{\mathbf{k}}^{(\pm)}$ and $\Delta_{\mathbf{k}}^{(\pm)}$
are projected onto the Fermi surface and depend only on the angle $\phi_{\mathbf{k}}$. The total DOS $\rho(\varepsilon)=\rho_{+}(\varepsilon)+\rho_{-}(\varepsilon)$, is then obtained as the sum of the separate contributions from the two bands. 

\subsection{Bulk properties: Bogoliubov bands and density of states}
\label{Bb_DOS}

In this subsection, we present examples of explicit predictions resulting from our model.
We focus on the structure of the Bogoliubov bands and the density of states in the presence of strong Rashba SOC, as also discussed above.
In Fig.~\ref{Fig:obs}(a) we plot the Bogoliubov bands resulting from one of the Rashba-split bands [$H^{(-)}$ as given in Eq.~(\ref{Eq:HRashba})], with the magnetic field set to zero. 
All other parameters are set to experimentally realistic values:
We fix the HH--LH splitting to be $-E_{\rm HL}= 100\, \Delta \approx 20$~meV (we work in units of $\Delta$, but have a value of $\Delta \approx 0.2$~meV in mind, which is realistic for Al), set the Rashba SOC energy to $\alpha_{\rm R} k_{\rm F}=10\,\Delta \approx 2$~meV, use $\zeta_F=0.35$ [see Eq.~(\ref{zetas})], set $\xi_{k_{\rm F},s}=0.05\,E_{\rm HL} \approx 1$~meV, and take the $g$-factors of the superconductor and 2DHG to be $g_s=-2$ and $\kappa = 3.4$, respectively.
Finally, we fix the parameters that characterize the superconductor--2DHG coupling to be $|\tau_{+-}| = 0.8$ and $\phi_t = \pi/4$ (corresponding to equal ``in-plane'' hopping strengths $t_{x}=t_{y}$).
The value of $t_z$ does not affect any physical properties in the current approximation, see the explicit expressions in Sec.~\ref{sec:Rashba_split}. 

The main qualitative difference of the resulting band structure compared to the case of a proximitized spin--orbit-coupled two-dimensional electron gas~\cite{Phan_2022,Babkin_2024} is the gap anisotropy along the Fermi surface we see in Fig.~\ref{Fig:obs}(a). 
This anisotropy can be quantified via $E_{u,\text{gap}}^{(\pm)} (\phi_{\mathbf{k}})$, with which we denote the $\phi_{\bf k}$-dependent distance from the minimum of the upper Bogoliubov band to the chemical potential, set to be at zero, see Fig.~\ref{Fig:obs}(a).
This gap function can be straightforwardly calculated from Eq.~(\ref{energy_spectrum}) as $E_{u,\text{gap}}^{(\pm)}\left(\phi_{\mathbf{k}}\right) = \min_{k}E_{u}^{\left(\pm\right)}\left(k,\phi_{\mathbf{k}}\right)$.

In Fig.~\ref{Fig:obs}(b) we plot the distance from the band bottom to the chemical potential $E_{u,\text{gap}}^{(-)}\left(\phi_{\mathbf{k}}\right)$ as a function of $\phi_{\bf k}$ (left), along with the single-band DOS $\rho^{(-)}(\varepsilon)$ as a function of (positive) energy $\varepsilon>0$ (right).
The upper plots have zero magnetic field $B=0$ and the lower plots use $B = 0.05\,\Delta$ with $\phi_B = 0$ (corresponding to a field of about $0.17$\,T). 
Extrema in the function $E_{u,\text{gap}}^{(\pm)}\left(\phi_{\mathbf{k}}\right)$ lead to singularities in the DOS (minima give discontinuities, maxima lead to logarithmic Van Hove singularities).
A finite magnetic field doubles the number of these singularities, due to the increased anisotropy of $E_{u,\text{gap}}^{(-)}\left(\phi_{\mathbf{k}}\right)$. 
In addition, the magnetic field suppresses the gap in the DOS and at sufficiently high fields the gap vanishes, where the 2DHG enters a gapless regime and  Bogoliubov Fermi surfaces emerge.

In Fig.~\ref{Fig:obs}(c) we show these Bogoliubov Fermi surfaces at a magnetic field $B = 0.7\,\Delta$ (corresponding to approximately $2.4$\,T).
The surfaces depicted in blue and orange correspond to the Rashba-split band $(-)$, while the red and green ones correspond to the band $(+)$. 
Comparing the bottom and top panels of Fig.~\ref{Fig:obs}(c) that correspond to a magnetic field along the $x$-axis and along the direction $\hat x + \hat y$, respectively, we see that $\phi_B$ qualitatively affects the structure of the Fermi surfaces by changing the number of pockets. 
This can be contrasted to the case of an \emph{electron} gas with strong Rashba SOC~\cite{Phan_2022,Babkin_2024}, where rotating the magnetic field only caused a trivial rotation of the Fermi surfaces in momentum space by the same angle. 
These nontrivial effects in the 2DHG arise  because the magnetic-field angle $\phi_B$ and the momentum angle $\phi_{\mathbf{k}}$ enter the Hamiltonian in a nontrivial way via the Zeeman term in Eq.~(\ref{Eq:Zeeman2x2}). 
The resulting interplay of different angular dependencies in Eq.~(\ref{Eq:Zeeman2x2}) leads to qualitative changes of the Bogoliubov bands depending on the orientation of the magnetic field. 
Additionally, the strength of the second term in Eq.~(\ref{Eq:Zeeman2x2}) differs between the two Rashba-split bands, explaining the distinct behavior for the inner (red and green) and outer (blue and orange) Fermi surfaces in Fig.~\ref{Fig:obs}(c). 

\subsection{Additional confinement: Effective $g$-tensor in proximitized quantum dots\label{Sec:g-factor}}

\begin{figure}[t]
\includegraphics[width=0.99\linewidth]{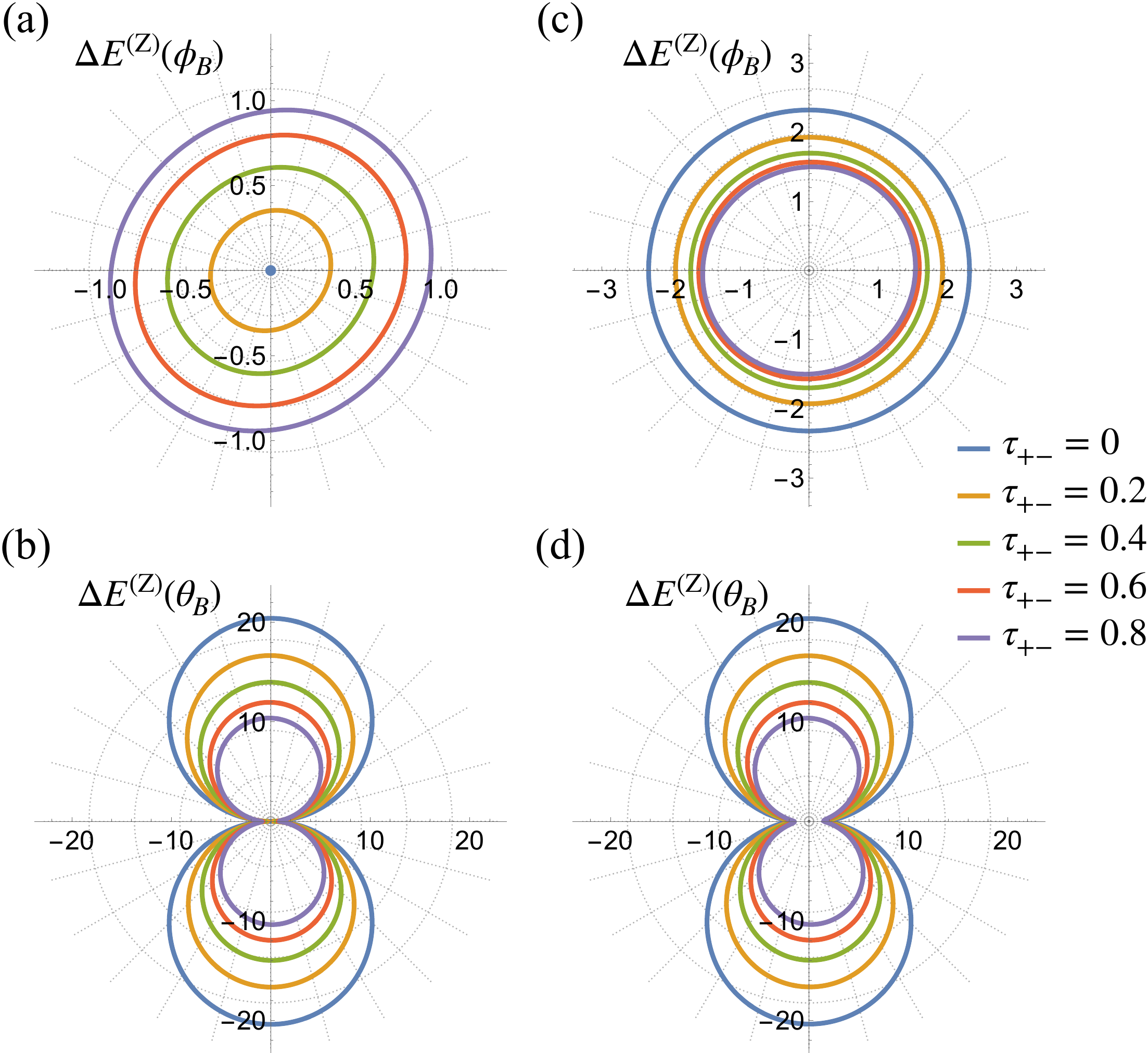}
\caption{\label{fig:gtensors} Angle-dependent Zeeman splittings in a proximitized quantum dot. 
We plot the splitting of the lowest-energy Kramers' pair per applied field ($\Delta E^{(Z)}/B$) visualized in polar plots as a function of the applied field direction for different coupling strengths to the superconductor ($\tau_{+-}$).
(a,b) Show the splitting in a symmetric dot ($d_y/d_x=1$, $d/d_x=0.15$) and (c,d) the splitting in an asymmetric dot ($d_y/d_x=0.3$, $d/d_x=0.15$).
In (a,c) the field is rotated in the $xy$-plane and in (b,d) in the $xz$-plane by angles $\phi_B$ and $\theta_B$, respectively. 
Additional parameters used are: $\kappa=3.4$, $g_s=-2$, $\phi_{t}=\pi/4$, $\xi_{k_{\rm F},s} = 0.05\, E_{\rm HL}$, and $|\tau_{z+}| = \tau_{+-}$.
For details on how the out-of-plane magnetic field is included, see App.~\ref{G_tensor_0D}.
}
\end{figure}

An in-plane magnetic field has to leading order no effect on the HH subspace, see Eq.~(\ref{H2DHG}) [up to terms cubic in angular momentum operators, which we neglect as discussed below Eq.~(\ref{zeeman-def})]. 
The effective HH Zeeman terms in Eqs.~(\ref{H11},\ref{H12}), and illustrated in Fig.~\ref{fig:diagrams_Z}, all arise due to coupling to the LH bands or the superconductor.
Additional confinement of the holes in the in-plane directions, e.g., forming a quasi-zero-dimensional quantum dot, yields bound states with discrete energy levels that in the absence of a magnetic field have a Kramers degeneracy.
The perturbative effect of a small magnetic field on such Kramers' pairs can be described by an effective spin-1/2 Hamiltonian
\begin{align}
 & \hat{H}^{(Z)}=-\frac{1}{2}\sum_{\alpha\beta}\hat{\sigma}_{\alpha}\hat{g}_{\alpha\beta}B_{\beta},\label{eq:genZ}
\end{align}
where $\alpha,\beta\in\left\{ x,y,z\right\} $, $\hat{\sigma}_{\alpha}$ is the vector of Pauli matrices, $B_{\beta}$ is the magnetic field, and $\hat{g}_{\alpha\beta}$ is the $3\times3$ effective $g$-tensor. 
In this subsection, we analyze how changing the size or shape of the quantum dot affects the $g$-tensor $\hat{g}_{\alpha\beta}$, through changes of the HH--LH coupling and the coupling to the superconductor.
From the effective Hamiltonian derived in Sec.~\ref{sec:effham} we can see that the dominating effect on the Zeeman terms is a renormalization of the $g$-factor due to leakage of the wave function into the superconductor, which has $g_s \approx - 2$; the results we find below thus resemble those for a hole quantum dot coupled to a simple normal metal.

In the simplest approach, the confinement can be modeled by projecting our Hamiltonian to the lowest-energy band in all directions. 
This means that we replace $k_{x,y}^2$ by their expectation values $\langle k_{x,y}^{2}\rangle = \pi^{2}/d_{x,y}^{2}$, where $d_{x,y}$ are the length scales of the in-plane confinement, and we ignore terms linear in $k_{x,y}$ which couple to excited states, similarly as we did in Sec.~\ref{sec:2dhg} to include the confinement along $\hat z$.
\edits{We emphasize that doing this in the effective HH Hamiltonian we derived in Sec.~\ref{sec:model} implicitly requires that $d_{x,y} \gg d$.}
Typically $d_{x,y}\sim 50$--$100$\,nm, which is indeed bigger than $d \sim 10$--20$\,$nm and yields a first excited quantum-dot level that is split off by a few meV.
For simplicity we will neglect all orbital effects below, which amounts to the assumption that $d_{x,y}, d \ll \sqrt{\hbar / eB}$, limiting the validity of our results to small magnetic fields.\footnote{Our analysis also ignores orbital effects in the superconductor which are beyond the scope of this work.}

In this limit, we can thus simply insert the following replacements into Eq.~(\ref{H11}) and (\ref{H12})
\begin{align}
    \langle\zeta_{\rm F}e^{-2i\phi_{\mathbf{k}}}\rangle &= \frac{\gamma_{\rm s}\pi^{2}}{m E_{\rm HL}}\left(\frac{1}{d_{x}^{2}}-\frac{1}{d_{y}^{2}}\right),\label{eq:<zetaFexp>}\\
    \langle\zeta_{\rm F}\cos(2\phi_{\mathbf{k}}-\tilde\phi)\rangle &= \frac{\gamma_{\rm s}\pi^{2}}{m E_{\rm HL}}\left(\frac{1}{d_{x}^{2}}-\frac{1}{d_{y}^{2}}\right)\cos\tilde \phi,\label{eq:<zetaFcos>}
\end{align}
where $\tilde \phi=\phi_{t}+\phi_{B}$ and $\langle{\cdot}\rangle$ denotes the projection onto the lowest transverse band as described above.
We identify the components of $\hat{g}_{\alpha\beta}$ by comparison of the resulting Hamiltonian with Eq.~(\ref{eq:genZ}). 
All components of the $g$-tensor we thus find, including out-of-plane components, are listed in App.~\ref{G_tensor_0D}. 

The resulting ``$g$-factor'', i.e., the Zeeman splitting per applied magnetic field 
$\Delta E^{(\rm Z)}/B$ of the confined HH-like spin as a function of applied magnetic field direction, is visualized in Fig.~\ref{fig:gtensors} for different coupling strengths $\tau_{+-}$ to the superconductor and different quantum-dot aspect ratios.
In Fig.~\ref{fig:gtensors}(a,c) the magnetic field is in-plane, the polar angle being $\phi_B$ and in Fig.~\ref{fig:gtensors}(b,d) we rotate the magnetic field in the $xz$-plane by the angle $\theta_B$, where $\theta_B=0$ corresponds to ${\bf B} \parallel \hat x$ and $\theta_B = \pi/2$ to ${\bf B} \parallel \hat z$.

The shape of the plots can be best understood by considering the special cases of a symmetric dot with no coupling to the superconductor first and then consecutively adding complexity.
For the symmetric dot ($d_{y}/d_{x}=1$) assumed in Fig.~\ref{fig:gtensors}(a), the Zeeman terms $\propto \zeta_{\rm F}$ (Z1 and Z4) vanish, see Eqs.~(\ref{eq:<zetaFexp>},\ref{eq:<zetaFcos>}). 
The remaining terms (Z2 and Z3) rely on coupling to the superconductor, explaining why the splitting vanishes for $\tau_{+-}=0$ [blue line in Fig.~\ref{fig:gtensors}(a)].
As the coupling to the superconductor is switched on, the term Z2 leads to an effective in-plane magnetic field rotated with respect to the applied magnetic field direction by $2\phi_{t}$. 
For the considered case of $t_{x}=t_{y}$, i.e., $\phi_{t}=\pi/4$, a field applied along the $x$-axis appears as an effective field along the $y$-axis and Z2 thus formally constitutes the components $g_{xy}=-g_{yx}$ in the $g$-tensor.
This term alone would yield an isotropic splitting, i.e., a constant $\Delta E_{\rm Z}$ as a function of $\phi_B$, that grows as $\tau_{+-}/(1+\tau_{+-})$ and asymptotically approaches $g_{s}$ as the coupling to the superconductor is increased.
The term Z3, however, represents an effective out-of-plane field resulting from the applied in-plane field, i.e., contributions to the components $g_{zx}$ and $g_{zy}$.
For $\phi_{t}=\pi/4$ we get $g_{zx}=g_{zy}$ and their contributions add up and are thus maximal for $B_x=B_y$, explaining the slight asymmetry in the shape of $\Delta E_{\rm Z}/B$ we observe in Fig.~\ref{fig:gtensors}(a).

In the case of an asymmetric dot, as assumed in Fig.~\ref{fig:gtensors}(b), we get finite contributions from the processes that involve the LH subspace (Z1 and Z4). 
The term Z1 corresponds to the components $g_{xx}=g_{yy}$, which in the uncoupled case $\tau_{+-} = 0$ are simply given by $g_{xx} = g_{yy} = -6\kappa\left\langle\zeta_{\rm F}e^{-2i\phi_{\mathbf{k}}}\right\rangle$, resulting in a symmetric Zeeman splitting $\Delta E_{\rm Z}$ (blue curve). 
For the large asymmetry ($d_{y}/d_{x}=0.3$) chosen here, the LH-induced components of $g_{xx}$ and $g_{yy}$ are dominant for all shown cases.
As the coupling to the superconductor is switched on, they are rescaled by the factor $1/(1+\tau_{+-})$ and further reduced by the proximity-induced components due to the opposite signs of $\kappa$ and $g_{s}$.
The already discussed contributions from Z2, i.e., $g_{xy}=-g_{yx}$, effectively counteract this reduction of the splitting, as, e.g., for $\phi=0$, $g_{xx}$ and $g_{yx}$ add up as $\Delta E^{(Z)}/B=\sqrt{g_{xx}^2+g_{yx}^2+g_{zx}^2}$. 
The effective out-of-plane field components due to a combination of Z3 and Z4 ($g_{zx}$ and $g_{zy}$) are now comparably small and the resulting asymmetry is very subtle.

For completeness we also show the effect of an out-of-plane magnetic field component in Fig.~\ref{fig:gtensors}(b,d) (see  App.~\ref{G_tensor_0D} for details) for the same two dot shapes. 
We rotate the field in the $xz$-plane and see that in this case the angle-dependence of the Zeeman splitting is strongly dominated by the large difference between the in-plane and out-of-plane $g$-factors in the HH subspace. As the coupling $\tau_{+-}$ is ramped up, the out-of-plane component is quickly reduced due to opposite signs of $g_s$ and $\kappa$.

\section{Summary and outlook
\label{sec:discuss}}

We have reconsidered the proximity-induced superconducting correlations in semiconductor hole gases due to the coupling to an $s$-wave superconductor.
In particular, we focused on the case where the hole gas is strongly confined in the direction perpendicular to the superconductor--semiconductor interface, thus forming a two-dimensional quantum well.
Because of atomic spin--orbit coupling, the relevant hole bands have total angular momentum $J=3/2$, with $j_z=\pm 3/2,\pm 1/2$, which seems to give rise to restrictive selection rules for the allowed Andreev processes at the interface. 
However, if one assumes that the interface to the superconductor does not respect microscopic rotational symmetry then the number of selection rules is reduced, allowing to describe the coupling between the superconductor and the semiconductor bands with three parameters, one for the coupling to each $p$-orbital. 
By integrating out the superconductor degrees of freedom, we derived the self-energy and a low-energy effective Hamiltonian for the hole system. 
It turns out that in the natural basis of the hole gas, both intraband and interband superconducting correlations are induced, of course still respecting the underlying $s$-wave pairing of the parent superconductor. 
When projecting to the lowest (heavy-hole) band in the semiconductor, the interband pairing terms are indeed important, inducing both effective $s$-wave and $d$-wave singlet pairings and, when combined with an interfacial Rashba spin--orbit interaction, also a triplet-like pairings.

\edits{
The specific dependence of our effective Hamiltonian on the material parameters allows to make a number of \emph{qualitative} predictions for the proximity effect. Specifically, the thickness $d$ of the 2DHG directly determines the HH--LH splitting $E_{\rm HL}$, which, in turn, controls the magnitude of all spin-mixing effects within the HH subband. Hence, larger thickness will thus enhance the terms corresponding to effective unconventional superconductivity ($d$-wave singlet and triplet pairings), but leave the conventional $s$-wave singlet term intact.
A higher chemical potential, controlled via, e.g., doping or external gating of the 2DHG,
increases the magnitude of the HH--LH mixing when approaching the light-hole band edge.
This can tune the magnitude of the effective $d$-wave and $p$-wave pairing terms.
An applied in-plane magnetic field nontrivially affects the band structure and density of states, effects we address in detail in the previous Section.
Finally, if one considers quantum dots, as we do in Sec.~\ref{Sec:g-factor}, then the strengths of the confinement along the $x$- and $y$-axes are additional input parameters, which can be easily tuned experimentally since this confinement is typically realized with top gates.
Depending on the symmetry of the confinement, one can obtain different forms for the $g$-tensor (or different dependencies of the Zeeman splitting on the direction of the magnetic field).
}

The effective Hamiltonian is a useful starting point for \edits{a \emph{quantitative}} understanding \edits{of}  a number of physical observables in proximitized hole gases that can be probed in experiment. 
We provided two examples: (1) we calculated the Bogoliubov Fermi surfaces and resulting density of states for the case of strong Rashba spin--orbit coupling and (2) we evaluated the proximity-modulated anisotropic $g$-tensor for quantum dots confined within the two-dimensional hole gas. 
We show how the full structure of the pairing matrix plays an important role.
The density of states is a standard quantity to measure in tunneling experiments~\cite{Ivar_Giaever1961}, and we highlight several proximity-induced features in the density of states that are sensitive to the direction and magnitude of an applied in-plane magnetic field.
An observation of the Bogoliubov Fermi surfaces that should appear in the regime of gapless induced superconductivity in the 2DHG is more challenging, as it requires momentum resolution~\cite{Zhu2021}.
However, we expect that the superfluid density measured via the kinetic inductance, similarly to how it was done for InAs~\cite{Phan_2022}, may offer an indirect insight into the structure of Bogoliubov Fermi surfaces and allow to fix the parameters of our model similar to the case of InAs~\cite{Babkin_2024}.
The dependence of the Zeeman splitting of a localized hole on the direction and magnitude of an externally applied magnetic field is straightforward to probe in experiment~\cite{Crippa2018,Jin2024}, and we show how the coupling to the superconductor modulates this dependence.
This provides another method to test the validity of our model and potentially estimate the underlying model parameters in real devices.

Next, we discuss limitations and possible extensions of our work.
\textit{Disorder:} 
In the derivation of the effective Hamiltonian we have assumed a clean interface which conserves in-plane crystal momentum.
In the presence of disorder, either in the superconductor itself or at the interface, this assumption is no longer valid.
Instead, the superconductor energy, which depends on momentum, should be replaced by an average over a range of momenta $\sim 1/\ell$ around the in-plane momentum of the carrier in the 2DHG (where $\ell$ is the elastic mean free path). 
However, this will not significantly change the main physical results because of the typically small semiconductor Fermi wave vector.
On the other hand, the dependence on the transverse quantization energy in the metal [see, e.g., Eq.~\eqref{hop_strength}] will lead to a strong dependence on the thickness of the metal film, which has been discussed earlier \cite{Mikkelsen2018}. 
For sufficiently wide films or strong disorder, this dependence becomes less pronounced again.
A detailed analysis of the importance of film thickness and, for example, surface roughness is called for.
\textit{Beyond simple $s$-wave superconductors:} 
The parent superconductor can have a more complex band structure or pairing symmetries than assumed in this work, which may influence the phenomenological parameters used to describe the interface. 
It would be interesting to understand, for example, the proximity effect resulting from coupling to $d$-wave superconductors, such as cuprates. 

Finally, we comment on other properties of germanium-based hole gases (or similar systems) for which our theory can be useful. 
We envision fruitful extensions of our theory to describe and explore, e.g., the superfluid density measurable via kinetic inductance~\cite{Phan_2022}, signatures of triplet pairing, induced topological superconductivity, the superconducting diode effect realized in the presence of a magnetic field~\cite{Yuan2022}, properties of proximitized germanium nanowires, Andreev processes in NS junctions, and SNS Josephson junctions. 

More broadly, our work describes the nontrivial interplay between orbital physics and the proximity effect due to coupling to a conventional $s$-wave superconductor. Our results suggest that proximitizing other families of (quasi-)two dimensional quantum materials, such as transition metal dichalcogenides~\cite{Trainer20,Ramezani21,Nieminen23},  with rich orbital physics could be a promising avenue for engineering novel superconducting states and hybrid quantum devices with tunable electronic and magnetic properties.  

\edits{
\emph{Note added:} After submitting our manuscript, a related study appeared~\cite{pino2025}, investigating a similar system and reaching largely comparable conclusions.} 

\begin{acknowledgments}
We acknowledge useful discussions with Georgios~Katsaros, Andrew~Higginbotham, and Oliver Schwarze.
This research was funded in part by the Austrian Science Fund (FWF) F 86, the European Research Council (Grant Agreement No.~856526), and by the DFG Collaborative Research Center (CRC) 183 Project No.~277101999. 
For the purpose of open access, the authors have applied a CC BY public copyright license to any Author Accepted Manuscript version arising from this submission.
\end{acknowledgments}

\appendix 

\section{Effective Hamiltonian\label{app:derive_H}}

In this appendix, we provide a detailed derivation of the effective Hamiltonian in Eq.~(\ref{eff_Hamilton}). 
In Sec.~\ref{app:self-en_der}, we present the derivation of the self-energy in the 2DHG, by explicitly calculating the path integral over the superconducting degrees of freedom.
Next, in Sec.~\ref{app:matrix_G}, we derive the explicit $8\times8$ matrix form of the Green function $G_{\rm 2DHG}$ for the proximitized 2DHG. 
Finally, in Sec.~\ref{app:Schr-W}, we apply the Schrieffer--Wolff transformation to $G_{\rm 2DHG}$ and project it onto the $j_z = \pm3/2$ subspace, deriving the effective $4\times4$ Green function $G$ and, as a result, the effective $4\times4$ Hamiltonian $H$.

\subsection{The self-energy in the 2DHG}
\label{app:self-en_der}
In this subsection of the appendix, we address the derivation of the self-energy in the 2DHG. 
We begin by rewriting the partition function from Eq.~(\ref{partition_all}),
explicitly separating the resonant transverse mode from other transverse modes in the superconducting film,
\begin{widetext}
\begin{align}
 & Z=\intop{\cal D}\left[\psi,\overline{\psi},\chi,\overline{\chi}\right]\exp\left[\sum_{n}\left(\frac{1}{2}\sum_{\mathbf{k},n_{z}\ne n_{z}^{(0)}}\overline{\Psi}_{\mathbf{k}n_{z}n}G_{{\rm SC},n_z}^{-1}\Psi_{\mathbf{k}n_{z}n}+\frac{1}{2}\sum_{\mathbf{k}}\left(\overline{\Psi}_{\mathbf{k}n}G_{\rm SC}^{-1}\Psi_{\mathbf{k}n}-\overline{\Psi}_{\mathbf{k}n}\Theta_{\mathbf{k}n}-\overline{\Theta}_{\mathbf{k}n}\Psi_{\mathbf{k}n}\right)\nonumber\right.\right.  \\
 & \hspace{16em} +\left.\left.\sum_{\mathbf{k},j=\pm\frac{3}{2},\pm\frac{1}{2}}i\epsilon_{n}\overline{\chi}_{\mathbf{k}nj}\chi_{\mathbf{k}nj}-{\cal H}_{\rm 2DHG}\left[\overline{\chi},\chi\right]\right)\right],\label{eq:partapp}
\end{align}
where we used the Green function in the superconductor, $G^{-1}_{{\rm SC},n_z}=i\epsilon_{n}-H_{\rm SC}$ ($G^{-1}_{SC}$ corresponds to the case $n_z=n^{(0)}_z$).
We use the bispinors
\begin{align*}
\Psi_{\mathbf{k}n_{z}n} = {} & {} \left(\psi_{\mathbf{k}n_{z}n\uparrow},\psi_{\mathbf{k}n_{z}n\downarrow},\overline{\psi}_{-\mathbf{k}n_{z}n\downarrow},-\overline{\psi}_{-\mathbf{k}n_{z},-n\uparrow}\right)^T,
\end{align*}
and the notation $\Psi_{\mathbf{k}n}=\Psi_{\mathbf{k}n_{z}^{(0)}n}$ to represent the resonant mode in the superconductor.
The bispinor $\Theta_{\mathbf{k}n}$ is
\begin{align*}
\Theta_{\mathbf{k}n} = {} & {} \left(\theta_{\mathbf{k}n,1},-\theta_{\mathbf{k}n,2},\overline{\theta}_{-\mathbf{k}-n,2},\overline{\theta}_{-\mathbf{k}-n,1}\right)^T,
\end{align*}
where $\theta_{\mathbf{k}n,1}=-t_{+}\chi_{\mathbf{k}n,3/2}+\frac{1}{\sqrt{3}}t_{-}\chi_{\mathbf{k}n,-1/2}+\sqrt{\frac{2}{3}}t_{z}\chi_{\mathbf{k}n,1/2}$ and
$\theta_{\mathbf{k}n,2}=\frac{1}{\sqrt{3}}t_{+}\chi_{\mathbf{k}n,1/2}-t_{-}\chi_{\mathbf{k}n,-3/2}-\sqrt{\frac{2}{3}}t_{z}\chi_{\mathbf{k}n,-1/2}$.
With this notation, the last two terms on the first line of (\ref{eq:partapp}) account for the hopping between the semiconductor and the superconductor since the hopping Hamiltonian can be expressed as ${\cal H}_{\rm hop} \left[\overline{\psi},\psi,\overline{\chi},\chi\right]=\frac{1}{2} \sum_{\mathbf{k}}\left(\overline{\Psi}_{\mathbf{k}n}\Theta_{\mathbf{k}n}+\overline{\Theta}_{\mathbf{k}n}\Psi_{\mathbf{k}n}\right)$.

At this point, the superconducting degrees of freedom corresponding to the modes $\Psi_{\mathbf{k}n_{z}n}$ with $n_{z}\ne n_{z}^{(0)}$ can be integrated out right away, resulting in an irrelevant constant prefactor, as they do not couple to the fields $\Theta_{\mathbf{k}n},\overline{\Theta}_{\mathbf{k}n}$.
To integrate out the degrees of freedom corresponding to the resonant mode $\Psi_{\mathbf{k}n}$, we complete the square as 
\begin{equation*}
\overline{\Psi}_{\mathbf{k}n}G_{\rm SC}^{-1}\Psi_{\mathbf{k}n}-\overline{\Psi}_{\mathbf{k}n}\Theta_{\mathbf{k}n}-\overline{\Theta}_{\mathbf{k}n}\Psi_{\mathbf{k}n}=
\left(\overline{\Psi}_{\mathbf{k}n}-\overline{\Theta}_{\mathbf{k}n}G_{\rm SC}\right)G_{\rm SC}^{-1}\left(\Psi_{\mathbf{k}n}-G_{\rm SC}\Theta_{\mathbf{k}n}\right)-
\overline{\Theta}_{\mathbf{k}n}G_{\rm SC}\Theta_{\mathbf{k}n},
\end{equation*}
and shift the integration variables $\psi,\overline{\psi}$.
Then one derives
\begin{align}
 Z=\intop{\cal D}\left[\psi,\overline{\psi},\chi,\overline{\chi}\right]\exp\left[ \sum_{n}\left( \frac{1}{2} \sum_{\mathbf{k}} \overline{\Psi}_{\mathbf{k}n} G_{\rm SC}^{-1} \Psi_{\mathbf{k}n} +
 \sum_{\mathbf{k}j}i\epsilon_{n}\overline{\chi}_{\mathbf{k}nj}\chi_{\mathbf{k}nj}-{\cal H}_{\rm 2DHG}\left[\overline{\chi},\chi\right]-\frac{1}{2}\sum_{\mathbf{k}}\overline{\Theta}_{\mathbf{k}n}G_{\rm SC}\Theta_{\mathbf{k}n}\right)\right],
\end{align}
where the summation over $j$ is again over the values $j = \pm3/2,\pm1/2$.
In this form, the integration over $\psi,\overline{\psi}$ is trivial
and results in an irrelevant constant prefactor, which is omitted
in the following.
The result is the final expression presented in Eq.~(\ref{partition_2DHG}).

This derivation of the self-energy can be easily applied to the case where all transverse modes contribute comparably to the hopping (which is especially relevant for thicker superconducting films). In this scenario, each transverse mode in the superconductor must be integrated out independently, in the same manner as the “resonant” mode discussed above. The self-energy then has an additional sum over the mode numbers $n_z$ in the superconductor, $\sum_{n_z}\overline{\Theta}_{\mathbf{k}nn_z}G_{{\rm SC},n_z}\Theta_{\mathbf{k}nn_z}$. The explicit dependence on $n_z$ is contained in the Green function $G_{{\rm SC},n_z}$ (specifically, in the dispersion $\xi_{\mathbf{k}n_z,s}$) and in $\Theta_{\mathbf{k}n}$, as the hopping constants $t_+,t_-,t_z$, in general, depend on $n_z$. 

\subsection{Matrix form of Green function for 2DHG}
\label{app:matrix_G}
In this subsection, we derive the explicit matrix form of the Green function $G_{\rm 2DHG}$ for the 2DHG. 
To achieve this, we introduce Nambu subspace, doubling the spinor space, to account for the induced pairing. 
Specifically, the doubled bispinor takes the form $(\check{{\cal X}}_{\mathbf{k}n})^T= ([{\cal X}_{\mathbf{k}n}^{(3/2)}]^T,[{\cal X}_{\mathbf{k}n}^{(1/2)}]^T )$, where ${\cal X}_{\mathbf{k}n}^{(j)}= (\chi_{\mathbf{k}n,j} , \chi_{\mathbf{k}n,-j} , \overline{\chi}_{-\mathbf{k},-n,-j} , -\overline{\chi}_{-\mathbf{k},-n,j} )^T$. 
This doubled bispinor is related to $\Theta_{\mathbf{k}n}$ by 
$\Theta_{\mathbf{k}n} = T\check{{\cal X}}_{\mathbf{k}}$,
where $T$ is the following matrix 
\begin{align}
T=\left(\begin{array}{cccccccc}
-t_{+} & 0 & 0 & 0 & \sqrt{\frac{2}{3}}t_{z} & \frac{t_{-}}{\sqrt{3}} & 0 & 0\\
0 & t_{-} & 0 & 0 & -\frac{t_{+} }{\sqrt{3}}& \sqrt{\frac{2}{3}}t_{z} & 0 & 0\\
0 & 0 & -t_{-}^{*} & 0 & 0 & 0 & -\sqrt{\frac{2}{3}}t_{z}^{*} & -\frac{t_{+}^{*}}{\sqrt{3}}\\
0 & 0 & 0 & t_{+}^{*} & 0 & 0 & \frac{t_{-}^{*}}{\sqrt{3}} & -\sqrt{\frac{2}{3}}t_{z}^{*}
\end{array}\right).
\end{align}
Thus, the self-energy $\Sigma$ in Eq.~(\ref{self-energ_Sigma}) takes the explicit matrix form
\begin{align}
\Sigma=T^{\dagger}G_{\rm SC}T.
\end{align}

Next, we proceed by explicitly calculating $G_{\rm SC}$,
\begin{align}
 G_{\rm SC}\approx\frac{1}{\left|\Delta\right|^{2}+\xi^2_{{k},s}-\varepsilon^{2}} \left(\begin{array}{cccc}
-\varepsilon-\xi_{{k},s} & -\frac{1}{2}g_s B_{-} & \Delta & 0\\
-\frac{1}{2}g_s B_{+} & -\varepsilon-\xi_{{k},s} & 0 & \Delta\\
\Delta^{*} & 0 & -\varepsilon+\xi_{{k},s} & -\frac{1}{2}g_s B_{-}\\
0 & \Delta^{*} & -\frac{1}{2}g_s B_{+} & -\varepsilon+\xi_{{k},s}
\end{array}\right).
\end{align}
Here, we neglect small corrections of the form $g_s B/\xi_{{k},s}$,
assuming that near the Fermi surface of the 2DHG, the superconducting dispersion $\xi_{{k},s}$ is large (e.g., in comparison to the Zeeman energy). 
In what follows, it is convenient to use the parameters $\tau_{\alpha\beta}$ introduced in Eq.~(\ref{hop_strength}).
Then, the self-energy can be written in the following form,
\begin{align}
 & \Sigma=\tau_{+-}\tilde{T}^{\dagger}\left(\begin{array}{cccc}
-\varepsilon-\xi_{{k},s} & -\frac{g_s B_{-}}{2} & \Delta & 0\\
-\frac{g_s B_{+}}{2} & -\varepsilon-\xi_{{k},s} & 0 & \Delta\\
\Delta^{*} & 0 & -\varepsilon+\xi_{{k},s} & -\frac{g_s B_{-}}{2}\\
0 & \Delta^{*} & -\frac{g_s B_{+}}{2} & -\varepsilon+\xi_{{k},s}
\end{array}\right)\tilde{T},
\label{explicit_sigma}
\end{align}
where $\tilde{T}=T/t_{+}$.

In the case when the resonant-mode approximation is not valid and all transverse modes contribute comparably, the self-energy becomes
\begin{align}
 \Sigma=\sum_{n_{z}}\frac{\left|t_{+,n_{z}}\right|^{2}}{\left|\Delta\right|^{2}+\xi^2_{{k}n_{z},s}-\varepsilon^{2}} \tilde{T}^{\dagger}
 \left(\begin{array}{cccc}
-\varepsilon-\xi_{{k}n_{z},s} & -\frac{g_sB_{-}}{2} & \Delta & 0\\
-\frac{g_sB_{+}}{2} & -\varepsilon-\xi_{{k}n_{z},s} & 0 & \Delta\\
\Delta^{*} & 0 & -\varepsilon+\xi_{{k}n_{z},s} & -\frac{g_sB_{-}}{2}\\
0 & \Delta^{*} & -\frac{g_sB_{+}}{2} & -\varepsilon+\xi_{{k}n_{z},s}
\end{array}\right)\tilde{T}.
\end{align}
\end{widetext}
Generalizing the parameters $\tau_{+-}$ and $\xi_{{k},s}$ as
\begin{eqnarray}
\tau_{+-}
&\rightarrow&
\sum_{n_{z}}\frac{t_{+,n_z}t_{-,n_z}}{\left|\Delta\right|^{2}+\xi^2_{{k}n_z,s}-\varepsilon^{2}},\\ 
\xi_{{k},s}\tau_{+-}
&\rightarrow&
\sum_{n_{z}}\xi_{{k}n_z,s}\frac{\left|t_{+,n_z}\right|^{2}}{\left|\Delta\right|^{2}+\xi^2_{{k}n_z,s}-\varepsilon^{2}}.
\end{eqnarray}
We see that $\Sigma$ retains the same form as in Eq.~(\ref{explicit_sigma}), with the only difference being the redefinition of the parameters $\tau_{+-}$ and $\xi_{{k},s}$.
Here, we implicitly assumed that the relative scaling between $t_{+}$, $t_{-}$ and $t_{z}$ remains unchanged as $n_{z}$ varies.

Finally, we extend the initial Hamiltonian $H_{\rm 2DHG}$ to Nambu space as well, resulting in the $8\times8$ matrix $H_{\rm 2DHG}^{8\times8}$.
In this space, the Green function takes the form
\begin{align}
 & G_{\rm 2DHG}^{-1}=\varepsilon-H_{\rm 2DHG}^{8\times8}-\Sigma,
\label{app:G2DHG}
\end{align}
where we moved from Matsubara frequencies to real frequencies.

\subsection{Spin-3/2 effective Hamiltonian}
\label{app:Schr-W}

We now derive the effective model for the $j_z = \pm3/2$ HH subspace.
To implement this, we project $G_{\rm 2DHG}^{-1}$ onto this subspace by applying a Schrieffer--Wolff transformation~\cite{winklerBook}.
The $j_z = \pm 3/2$ subspace corresponds to the upper left $4\times4$ block
of Eq.~(\ref{app:G2DHG}).
We perturbatively rotate the matrix $G_{\rm 2DHG}^{-1}$ to eliminate the off-diagonal $4\times4$ blocks up to leading order in $1/E_{\rm HL}$,
\begin{equation}\label{Eq:SW}
 \tilde{G}_{\rm 2DHG}^{-1}=e^{-S}G_{\rm 2DHG}^{-1}e^{S},
\end{equation}
introducing the unitary transformation $e^{S}$.

In order to find $S$ and the effective $\tilde{G}_{\rm 2DHG}^{-1}$, we decompose $G_{\rm 2DHG}^{-1}$ as $G_{\rm 2DHG}^{-1}=A+B$, where $B$ contains two $4\times4$ off-diagonal blocks and $A$ contains the rest. 
Expanding Eq.~(\ref{Eq:SW}) to second order in $S$, we obtain
\begin{equation}
  \tilde{G}_{\rm 2DHG}^{-1}\approx A+ B+\left[A,S\right]+\left[B,S\right]+\frac{1}{2}\left[\left[A,S\right],S\right].
\end{equation}
Next, we explicitly separate $E_{\rm HL}$ from $A$,
$$A=A_{0}+\left(A-A_{0}\right),$$
where $A_{0}$ contains all terms involving $E_{\rm HL}$: $A_{0}=\text{diag}\left(0,0,0,0,-E_{\rm HL},-E_{\rm HL},E_{\rm HL},E_{\rm HL}\right)$.
To eliminate the off-diagonal blocks, we impose $B+\left[A,S\right]=0$ which, to leading order, is equivalent to $B=-\left[A_{0},S\right]$.
The rotated Green function then takes the form
\begin{equation}\label{Eq:G-tilde}
  \tilde{G}_{\rm 2DHG}^{-1}\approx A+\frac{1}{2}\left[B,S\right].
\end{equation}
In this approximation, we neglect terms in the diagonal blocks of order $1/E_{\rm HL}^{2}$ and terms in the off-diagonal blocks of order $1/E_{\rm HL}$.
Now, the upper-left $4\times4$ block of Eq.~(\ref{Eq:G-tilde}) can be treated as the effective inverse Green function for the HH $j_z = \pm 3/2$ subspace.
We will denote this Green function as $G$ hereafter.

We find that the Green function $G$ can be written as
\begin{widetext}
\begin{align}
G^{-1}=\left(\begin{array}{cccc}
\left[G^{-1}\right]_{11} & \left[G^{-1}\right]_{12} & \left[G^{-1}\right]_{13} & \left[G^{-1}\right]_{14}\\
\left(\left[G^{-1}\right]_{12}\right)^{*} & \left[G^{-1}\right]_{11}\big|_{\mathbf{k}\rightarrow-\mathbf{k},\mathbf{B}\rightarrow-\mathbf{B}} & \left(\left[G^{-1}\right]_{14}\right)^{*} & \left[G^{-1}\right]_{13}\big|_{\mathbf{k}\rightarrow-\mathbf{k}}\\
\left(\left[G^{-1}\right]_{13}\right)^{*} & \left[G^{-1}\right]_{14} & -\left[G^{-1}\right]_{11}\big|_{\mathbf{B}\rightarrow-\mathbf{B},\varepsilon\rightarrow-\varepsilon} & \left[G^{-1}\right]_{12}\big|_{\mathbf{k}\rightarrow-\mathbf{k}}\\
\left(\left[G^{-1}\right]_{14}\right)^{*} & \big(\left[G^{-1}\right]_{13}\big|_{\mathbf{k}\rightarrow-\mathbf{k}}\big)^{*} & \big(\left[G^{-1}\right]_{12}\big|_{\mathbf{k}\rightarrow-\mathbf{k}}\big)^{*} & -\left[G^{-1}\right]_{11}\big|_{\mathbf{k}\rightarrow-\mathbf{k},\varepsilon\rightarrow-\varepsilon}
\end{array}\right),
\end{align}
showing that most of the matrix elements are related to each other, which results in $G$ being fully defined by the four matrix elements $[G^{-1}]_{11},[G^{-1}]_{12},[G^{-1}]_{13},[G^{-1}]_{14}$.

Explicitly, we find for the matrix elements
\begin{align}
 [G^{-1}]_{11}= {} & {} \left(1+\tau_{+-}\right)\varepsilon - \xi_{{k}}+\frac{\alpha_{\rm R}k\left(6\kappa+\tau_{+-}g_s\right)  B }{E_{\rm HL}}\sin\left(\phi_{\mathbf{k}}-\phi_{B}\right)-\frac{\left|\tau_{z+}\right|}{\sqrt 2}\frac{\gamma_{s}k^{2}}{m E_{\rm HL}}g_s B \cos\left(2\phi_{\mathbf{k}}-\phi_{t}-\phi_{B}\right) \nonumber \\
 {} & {}-2\sqrt{2}\left|\tau_{z+}\right|\left(\xi_{{k},s} +\varepsilon\right) \left[ \frac{\alpha_{\rm R}k}{E_{\rm HL}} \sin\left(\phi_{\mathbf{k}}-\phi_{t}\right)+\frac{\kappa B}{E_{\rm HL}}\cos\left(\phi_{B}-\phi_{t}\right) \right] -\tau_{+-}\left(\xi_{{k},s}+\varepsilon\right)\frac{\gamma_{s}k^{2}}{m E_{\rm HL}}\cos\left[2\left(\phi_{\mathbf{k}}-\phi_{t}\right)\right],
\label{G11inv}\\
 [G^{-1}]_{12} = {} & {} 3\frac{\gamma_{s}k^{2}}{mE_{\rm HL}} e^{-2i\phi_{\mathbf{k}}} \left[i\alpha_{\rm R}ke^{-i\phi_{\mathbf{k}}} + \left(\kappa+\frac{1}{6}\tau_{+-}g_s\right) B e^{-i\phi_{B}}\right]
 \nonumber\\  &
 - 2i \tau_{+-} \frac{\xi_{{k},s}+\varepsilon}{E_{\rm HL}}\alpha_{\rm R} ke^{-i\left(\phi_{\mathbf{k}}+2\phi_{t}\right)} - 
 \frac{\tau_{+-}}{2}  g_sB e^{-i(\phi_{B}+2\phi_{t})},\label{eq:g12}\\
 [G^{-1}]_{13}= {} & {} -\Delta\tau_{+-}\left(1-\frac{\gamma_{s}k^{2}}{m E_{\rm HL}}\cos\left[2\left(\phi_{\mathbf{k}}-\phi_{t}\right)\right]\right)+2\sqrt{2}\Delta\left|\tau_{z+}\right|\left[\frac{\alpha_{\rm R}k}{E_{\rm HL}}\sin\left(\phi_{\mathbf{k}}-\phi_{t}\right)-i\frac{\kappa B}{E_{\rm HL}}\sin\left(\phi_{B}-\phi_{t}\right)\right],\\
 [G^{-1}]_{14} = {} & {} i\Delta\tau_{+-}\frac{2\alpha_{\rm R}k}{E_{\rm HL}}e^{-i\left(\phi_{\mathbf{k}}+2\phi_{t}\right)},
\end{align}
using the same notations as in the main text and introducing
\begin{align}
&\xi_{{k}}=\frac{k^{2}\left(\gamma_{1}+\gamma_{s}\right)}{2m}-\mu_h-\tau_{+-}\xi_{{k},s}-\frac{3\gamma_{s}^{2}k^{4}}{4E_{\rm HL} m^{2}}-\frac{3\alpha_{\rm R}^{2}k^{2}}{E_{\rm HL}},
\label{xi_all_comp}
\end{align}
\end{widetext}
where the last three terms in $\xi_{{k}}$ can be neglected since they result in a shift of chemical potential and small renormalization of the effective mass. 
We omitted contributions in $[G^{-1}]_{12}$ that result in a small renormalization of the last term in Eq.~(\ref{eq:g12}).
In what follows, we will also neglect the dependence on $\varepsilon$ in $[G^{-1}]_{12}$, since it is suppressed by $E_{\rm HL}$.
In the expression for $[G^{-1}]_{13}$, we neglected the term that has the same form as $\Delta\tau_{+-}$ but is suppressed by the small factor $\xi_{k,s}/E_{\rm HL}$. 
Hereafter, we will also neglect the weak $\varepsilon$-dependence of $\tau_{+-}$, since it is suppressed by $\varepsilon/\xi_{{k},s} \ll 1$.
Additionally, we will neglect the dependence
on $\varepsilon$ of all terms on the right-hand side of Eq.~(\ref{G11inv}), as it is suppressed by $E_{\rm HL}$, except for the first term.
Besides, we can take $\xi_{{k},s}$ at the Fermi surfaces of the 2DHG as only the vicinity of the Fermi surface is relevant for the observables studied in this paper (such as the DOS and the $g$-factor).
Any corrections arising from the momentum-dependence of $\xi_{{k},s}$ away from the Fermi surface are much smaller compared to its value at the Fermi surface, as $\xi_{{k},s}$ is much bigger than the relevant low-energy scales (such as $\Delta$ and $B$). 
 
Finally, this yields a Green function $G$ of the form
\begin{align}
G^{-1}=\left(1+\tau_{+-}\right)\varepsilon-H,
\end{align}
where $H$ does not depend on $\varepsilon$ and serves as the effective Hamiltonian presented in the main text [see Eq.~(\ref{eff_Hamilton})].

\section{Rashba-split bands \label{app:Rashba_split}}

In this appendix, we analyze the contribution from Rashba SOC to the effective Hamiltonian in Eq.~(\ref{eff_Hamilton}), explicitly separating the Rashba-split bands and neglecting the interband coupling.

The leading contribution from Rashba SOC is represented by the term R1 in Eq.~(\ref{H12}). 
By applying the rotation $H\rightarrow U^{\dagger}HU$ with the matrix
\begin{align}
 & U=\frac{1}{\sqrt{2}}\left(\begin{array}{cccc}
-ie^{-3i\phi_{\mathbf{k}}} & 0 & ie^{-3i\phi_{\mathbf{k}}} & 0\\
1 & 0 & 1 & 0\\
0 & -ie^{-3i\phi_{\mathbf{k}}} & 0 & ie^{-3i\phi_{\mathbf{k}}}\\
0 & 1 & 0 & 1
\end{array}\right),
\end{align}
we diagonalize the contribution R1, thus explicitly separating the
Rashba-split bands. The rotated Hamiltonian is expressed as
\begin{align}
 & U^{\dagger}HU=\left(\begin{array}{cc}
H^{(-)} & V^{(\text{off})}\\
(V^{(\text{off})})^{\dagger} & H^{(+)}
\end{array}\right),
\end{align}
where the diagonal blocks $H^{(\pm)}$ are defined in Eq.~(\ref{H_simp}) and the components of the off-diagonal are
\begin{widetext}
\begin{subequations}
\begin{eqnarray} V_{12}^{(\text{off})} = V_{21}^{(\text{off})}\Big|_{\Delta\rightarrow \Delta^*} & = & 
2\sqrt{2}\Delta \left|\tau_{z+}\right| \frac{\alpha_{\rm R}k}{E_{\rm HL}}\sin\left(\phi_{\mathbf{k}}-\phi_{t}\right)
+ 2i\Delta\tau_{+-} \frac{\alpha_{\rm R}k}{E_{\rm HL}}\sin\left[2(\phi_{\mathbf{k}}-\phi_{t})\right],\\
V_{11}^{(\text{off})}=V_{22}^{(\text{off})}\Big|_{\alpha_{\rm R}\rightarrow-\alpha_{\rm R}} & = & -2\sqrt{2}\left|\tau_{z+}\right|\frac{\xi_{{k},s}}{E_{\rm HL}}\left[\alpha_{\rm R}k\sin\left(\phi_{\mathbf{k}}-\phi_{t}\right)+\kappa B\cos\left(\phi_{B}-\phi_{t}\right)\right]-2i\tau_{+-}\alpha_{\rm R}k\frac{\xi_{{k},s}}{E_{\rm HL}}\sin\left[2\left(\phi_{\mathbf{k}}-\phi_{t}\right)\right]\nonumber\\
 & & -\frac{1}{\sqrt 2}g_s B\left|\tau_{z+}\right| \frac{\gamma_s k^2 }{m E_{\rm HL}} \cos\left(\phi_{B}-2\phi_{\mathbf{k}}+\phi_{t}\right)+\frac{i}{2}g_s B\tau_{+-}\cos\left(\phi_{B}-3\phi_{\mathbf{k}}+2\phi_{t}\right)
 \nonumber\\
 & & -\frac{i}{2}\left(6\kappa+ g_s\tau_{+-}\right)B\frac{\gamma_s k^2 }{m E_{\rm HL}}\cos\left(\phi_{B}-\phi_{\mathbf{k}}\right).
\end{eqnarray}
\end{subequations}
\end{widetext}
Assuming that the magnetic field is small enough such that $B\lesssim\sqrt{\zeta_{\rm F} \xi_{{k},s}/E_{\rm HL}}\alpha_{\rm R}k_{\rm F}$, where $\zeta_{\rm F}$ is defined in (\ref{zetas}), we can neglect the off-diagonal blocks $V^{(\text{off})}$, which describe the interband coupling between the Rashba-split bands.
This restriction can be straightforwardly derived by applying the Schrieffer--Wolff transformation (as outlined in App.~\ref{app:Schr-W}) and requiring the smallness of the emergent corrections to the diagonal blocks. 
When making this estimation, we used the following realistic
assumptions for the parameters: $|\Delta|\ll\xi_{{k},s}\ll\alpha_{\rm R}k_{\rm F}\ll\zeta_{\rm F}E_{\rm HL}\ll E_{\rm HL}$, $B\ll E_{\rm HL} \zeta^2_{\rm F}$,
$\tau_{+-}\sim1$, $\left|\tau_{z+}\right|\sim1$, $g_s\sim1$. 
As a result, the simplified Hamiltonian takes the form shown in Eq.~(\ref{H_simp}).

\section{Zeeman effect in zero-dimensional case \label{G_tensor_0D}}

In this appendix we derive the effective Zeeman terms for an out-of-plane field component. 
These results are used to calculate all $g$-tensor components of a bound state in an effectively zero-dimensional quantum dot. 
In our derivations below, we ignore orbital effects in the quantum dot, which is justified when the electronic wave function is confined to length scales smaller than $\sqrt{\hbar/eB}$. 

The presence of a finite $z$-component of the magnetic field, $B_{z}$, results in the following additional contributions to the Hamiltonian of the superconductor and 2DHG, respectively: $\delta H_{\rm SC}=-\frac{1}{2}g_{s}B_{z}\sigma_{z}$,
$\delta H_{\rm 2DHG}=-2\kappa B_{z}J_{z}$. Then, the Green function
of the superconductor takes the following form
\begin{widetext}
\begin{align}
 & G_{\rm SC}\approx\frac{1}{\left|\Delta\right|^{2}+\xi^2_{{k},s}-\varepsilon^{2}}\left(\begin{array}{cccc}
-\varepsilon-\xi_{{k},s}-\tfrac{1}{2}g_{s}B_{z} & -\tfrac{1}{2}g_{s}B_{-} & \Delta & 0\\
-\tfrac{1}{2}g_{s}B_{+} & -\varepsilon-\xi_{{k},s}+\tfrac{1}{2}g_{s}B_{z} & 0 & \Delta\\
\Delta^{*} & 0 & -\varepsilon+\xi_{{k},s}-\tfrac{1}{2}g_{s}B_{z} & -\tfrac{1}{2}g_{s}B_{-}\\
0 & \Delta^{*} & -\tfrac{1}{2}g_{s}B_{+} & -\varepsilon+\xi_{{k},s}+\tfrac{1}{2}g_{s}B_{z}
\end{array}\right),
\end{align}
where $B_{\pm}=B_{\parallel}e^{\pm i\phi_{B}}$ in terms of the in-plane
component of magnetic field $B_\parallel$. 
Proceeding as in App.~\ref{app:derive_H}, we derive the effective Green function $G$ of the proximitized 2DHG for the HH $j_z = \pm \frac{3}{2}$ subspace. 
Since we are now focusing on the Zeeman effect, we restrict our analysis to the upper left $2\times2$ block of the inverse Green function, $G^{-1}$. The matrix elements in this block acquire additional contributions due to the presence of $B_{z}$, which are given by
\begin{eqnarray}\label{Eq:H11Bz}
  \delta[G^{-1}]_{11}&=&
  \frac{1}{2}\left(6\kappa +g_s \tau_{+-} \right) B_{z}-\frac{1}{2}g_{s}B_{z}\left(\tau_{+-}\zeta_{\rm F}\cos\left[2\left(\phi_{\mathbf{k}}-\phi_{t}\right)\right]-\frac{1}{3}\left[2\left|\tau_{z+}\right|^{2}+\tau_{+-}^{2}\right]\left[\zeta_{s}+\frac{2\varepsilon}{E_{{\rm HL}}}\right]\right)
  \nonumber\\
 & & -\frac{1}{\sqrt{2}} g_{s} B_{z} \left|\tau_{z+}\right|\zeta_{\rm R}\sin\left(\phi_{\mathbf{k}}-\phi_{t}\right),
 \\
 \delta[G^{-1}]_{12}&=&
 -\frac{1}{\sqrt{2}} g_{s}B_{z} \left|\tau_{z+}\right|\zeta_{\rm F}e^{-i\left(2\phi_{\mathbf{k}}+\phi_{t}\right)}.
\end{eqnarray}
\end{widetext}
In the matrix element $\delta[G^{-1}]_{11}$ we omitted contributions quadratic in magnetic field, which result in a shift of the chemical potential. 
The two other matrix elements are defined in the following way: $\delta[G^{-1}]_{22} = \delta[G^{-1}]_{11}|_{\mathbf{k}\rightarrow-\mathbf{k},\mathbf{B}\rightarrow-\mathbf{B}}$ and $\delta[G^{-1}]_{21}=\left(\delta[G^{-1}]_{12}\right)^{*}$. In what follows, we will neglect the weak $\varepsilon$-dependence of $\delta [G^{-1}]_{11}$ and $\delta [G^{-1}]_{22}$ since it is suppressed by the energy scale of HH--LH splitting $E_{\rm HL}$, unlike the leading $\varepsilon$-dependence
in Eq.~(\ref{G11inv}). 
We further note that the last term of Eq.~(\ref{Eq:H11Bz}) results in the shift of Fermi surface, and does therefore not contribute to the Zeeman effect.
This term can thus be omitted. 

Finally, we collect all Zeeman contributions from the in-plane and out-of-plane field into a contribution to the effective Hamiltonian,
\begin{eqnarray}
    H_{11}^{Z} &=& {Z_3}+{Z_4}+{Z_5}+{Z_6},\\
    H_{12}^{Z} &=& {Z_1}+{Z_2}+{Z_7},
\end{eqnarray}
where we use the same block structure as in Eqs.~(\ref{H11},\ref{H12}) in the main text.
The terms $Z_1$--$Z_4$ are defined in Eqs.~(\ref{H12},\ref{H11}), where $B$ should now be replaced by $B_{||}$. 
The remaining terms $Z_5$--$Z_7$ are contributions to Zeeman energy from the out-of-plane component of magnetic field, and read explicitly as
\begingroup
\allowdisplaybreaks
\begin{align}
 & Z_{5}=-\frac{1}{2}\left(6\kappa+g_{s}\tau_{+-} \right)B_{z} \nonumber\\
 & \hspace{3em} -\frac{1}{6}g_{s}\zeta_s
 (2\left|\tau_{z+}\right|^2+\tau^2_{+-})B_{z},\\
 & Z_{6}=\frac{1}{2}g_{s}\tau_{+-}\zeta_{\rm F}\cos\left[2\left(\phi_{\mathbf{k}}-\phi_{t}\right)\right]B_{z},\\
 & Z_{7}=\frac{\sqrt{2}}{2}g_{s}\left|\tau_{z+}\right| 
 \zeta_{\rm F}e^{-i\left(2\phi_{\mathbf{k}}+\phi_{t}\right)}B_{z}.
\end{align}
\endgroup

Replacing $k_{x,y}$ and $k_{x,y}^{2}$ by their expectation values as discussed in Sec.~\ref{Sec:g-factor}, we find all components of the $g$-tensor as defined in Eq.~(\ref{eq:genZ}):
\begin{subequations}
\begin{align}    
    \begin{split}
    g_{xx}=&g_{yy}=-\left(6\kappa+g_s\tau_{+-}\right)\left\langle\zeta_{\rm F}e^{-2i\phi_{\mathbf{k}}}\right\rangle\\
    &+g_s\tau_{+-}\cos2\phi_t,
    \end{split}\\
    g_{xy}=&-g_{yx}=-g_s\tau_{+-} \sin2\phi_t,
    \\
    g_{zx}=&2\sqrt{2}|\tau_{z+}|\left(\kappa\zeta_s - \frac{g_s}{2}\left\langle\zeta_{\rm F}e^{-2i\phi_{\mathbf{k}}}\right\rangle\right)\cos\phi_t,
    \\
    g_{zy}=&2\sqrt{2}|\tau_{z+}|\left(\kappa\zeta_s+ \frac{g_s}{2}\left\langle\zeta_{\rm F}e^{-2i\phi_{\mathbf{k}}}\right\rangle\right)\sin\phi_t,
    \\
    \begin{split}
    g_{zz}=& -6\kappa-g_s\tau_{+-}-\frac{g_{s}}{6}\left(2\left|\tau_{z+}\right|^2+\tau^2_{+-}\right)\zeta_s\\
    &+g_s\tau_{+-}\left\langle\zeta_{\rm F}e^{-2i\phi_{\mathbf{k}}}\right\rangle\cos2\phi_{t},
    \end{split}
    \\
    g_{xz}=&\sqrt{2}g_{s}|\tau_{z+}|\left\langle\zeta_{\rm F}e^{-2i\phi_{\mathbf{k}}}\right\rangle\cos \phi_t,
    \\
    g_{yz}=&\sqrt{2}g_{s}|\tau_{z+}|\left\langle\zeta_{\rm F}e^{-2i\phi_{\mathbf{k}}}\right\rangle\sin \phi_t,
\end{align}
\end{subequations}
where $\langle{\cdot}\rangle$ denotes the projection onto the lowest transverse band, as defined in Eq.~(\ref{eq:<zetaFexp>}). Note that to get the physical splittings in Fig.~\ref{fig:gtensors} we still need to include the renormalization factor $1/(1+\tau_{+-})$.

\bibliography{bib_germanium}

\end{document}